\documentclass[%
 reprint,
 amsmath,amssymb,
 aps,
pre,
]{revtex4-1}

\usepackage{dsfont}
\usepackage{graphicx}% Include figure files
\usepackage{dcolumn}% Align table columns on decimal point
\usepackage{bm}% bold math
\usepackage{hyperref}% add hypertext capabilities
\usepackage{subfig}
\usepackage{microtype}
\usepackage{tikz}
\usepackage{pgfplots, pgfplotstable}
\usepackage{tkz-graph}
\usetikzlibrary{arrows}
\usetikzlibrary{matrix}
\usetikzlibrary{patterns}
\usetikzlibrary{shapes}
\usetikzlibrary{shapes.multipart}
\usepgfplotslibrary{groupplots}
\usetikzlibrary{pgfplots.groupplots}

\definecolor{orange}{rgb}{1,0.3,0.1}
\definecolor{lila}{rgb}{0.9,0,0.3}
\definecolor{green}{rgb}{0,0.6,0}
\definecolor{hell}{rgb}{0.6,0.6,0}
\definecolor{white}{rgb}{1.0,1.0,1.0}

\newcommand{\black}[1]{\textcolor{black}{#1}}

\renewcommand{\vec}{\boldsymbol}

\newcommand{\JM}[1]{\black{#1}}

\graphicspath{{./Fig/}}

\begin{document}

\preprint{APS/123-QED}

\title{Social Influence with Recurrent Mobility with multiple options}% Force line breaks with \\

\author{J\'er\^ome Michaud}
 %\altaffiliation[Also at ]{Physics Department, University of Edinburgh.}%Lines break automatically or can be forced with \\
 \email{jerome.michaud@soc.uu.se}
\affiliation{%
Department of Sociology \& Departement of physics and astronomy, University of Uppsala
}%


\author{Attila Szilva}
\affiliation{
Departement of physics and astronomy, University of Uppsala
}%
%\affiliation{
%\affiliation{%


\date{\today}% It is always \today, today,
\begin{abstract}
In this paper, we discuss the possible generalizations of the Social Influence with Recurrent Mobility (SIRM) model developed in Phys. Rev. Lett. {\bf 112}, 158701 (2014). Although the SIRM model worked approximately satisfying when US election was modelled, it has its limits: it has been developed only for two-party systems and can lead to unphysical behaviour when one of the parties has extreme vote share close to 0 or 1. We propose here generalizations to the SIRM model by its extension for multi-party systems that are mathematically well-posed in case of extreme vote shares, too, by handling the noise term in a different way. In addition, we show that our method opens new applications for the study of elections by using a new calibration procedure, and makes possible to analyse the influence of the ``free will" (creating a new party) and other local effects for different commuting network topologies.

\end{abstract}

\pacs{Valid PACS appear here}% PACS, the Physics and Astronomy
\maketitle

\if 0
\section{Structure of the paper}
\begin{itemize}
\item Introduction: General motivation + context of the paper
\item Generalized Voter Model with Recurrent Mobility (GVMRM)
\begin{itemize}
\item Recurrent Mobility structure (Introduce notation, topology and vote shares)
\item Dynamics: Introduce the general formalism, that is $R^{kk'}_{ij}$
\end{itemize}
%\item SIRM in general: Present the general formalism, that is $R^{kk'}_{ij}$
\item Present the SIRM model of the Mallorca group and explain why it does not fit in the general framework. Illustrate the problems that this implies
\item How to put noise on  $R^{kk'}_{ij}$? Uses the Dirichlet distribution to create noise. Show that the SIRM model can be recovered under appropriate assumptions form the generalized model.
\item Apply the model on test cases. Fully connected network, look for stationarity in the results by changing $\gamma$ and $D$.
\item Conclusion
\item Supplementary material: Details about implementation of the model. (Jupyter notebook with the code?)
\end{itemize}

\fi

\section{Introduction}\label{sec:Intro}
Statistical physics models to study social systems have been used for quite some time now, see, for example, the review paper \cite{RevModPhys.81.591}. One of the subdomain in which statistical physics models have been used is opinion dynamics and voting behaviour. In this context, the Voter Model (VM) introduced by \cite{doi:10.1093/biomet/60.3.581} and named ``Voter Model" in \cite{holley1975} had a huge influence and many scholars have explored its dynamics in various settings. The VM has been studied on regular lattices \cite{PhysRevE.53.R3009,PhysRevA.45.1067}, where an exact solution exists, or random networks \cite{wu2004information,PhysRevE.72.036132,PhysRevLett.94.178701,sood2008voter}, and various modifications of it have been proposed to make it more realistic \cite{sire1995coarsening,PhysRevLett.91.028701,1742-5468-2007-08-P08029,vazquez2003constrained,lambiotte2007dynamics,dall2007effective}. However, most of these works have focused on the VM from the mathematical point of view and did not try to compare it with actual elections data. Therefore, the question ``Is the Voter Model a Model for Voters?" still needs to be addressed, especially in extreme cases when the vote share is close to $0$ or $1$.

A first step towards answering this question has been proposed by %\AS{Ref.} Removed, because I do not use this later on
 \cite{PhysRevLett.112.158701}. In this paper, the authors proposed a modification of the VM that they called \emph{Social Influence with Recurrent Mobility}, abbreviated as SIRM. In this model, the network underlying the VM is inferred from the commuting pattern of voters. A commuter being an individual living in one place and working in another. Using census data, it is possible to construct the commuting network of the population, which is defined at the community level rather than at the individual level, assuming that all members of a community are connected together. In the application to the US election \cite{PhysRevLett.112.158701}, counties were chosen as the relevant communities and two counties are linked by an edge if at least one individual lives in one the counties and works in the other. The edge is weighted by the total number of commuters. Furthermore, two parameters have been added to the standard VM. The first parameter, $\alpha$, controls the probability that an individual interacts where she lives or where she works. The second parameter, $D$, is a noise parameter accounting for any other factors that can influence the change of opinion, such as free will or mass media. Using this model, it was possible to recover some statistical features of US presidential election\JM{s}, such as the shape of the distribution of vote shares around the average and the logarithmic decay of the spatial correlation\JM{s} of vote shares \cite{PhysRevLett.112.158701}.

As we will show, it turns out that the SIRM model is not mathematically well-posed and that it can predict unphysical behaviour of the population if one of the party has a vote share that is either close to $0$ or to $1$. In the case of the US election, the importance of this issue is small, since the two parties considered roughly share the votes evenly. However, if we would like to generalize the SIRM model to an arbitrary number of parties, this issue becomes critical to avoid number of voters to become negative.

In this paper, we study the SIRM model and its extension from an abstract point of view. We will not apply it to real data here, instead, we would like to point out that the structure implied by the commuting network is %\JM{\it Removed We would like to point out the structure implied by the commuting network instead that is}
 interesting in itself and extends the usual topological discussion of the dynamics of the VM. For instance, the commuting network is a weighted directed graph with self-loops and the influence of its topology on SIRM like models is yet to be studied. In this paper, we restrict ourselves to some simple commuting networks, since a detailed study of the influence of the topology of the commuting network could be the subject of an entire paper. Apart from the topology, the influence of the initial voting distribution is likely to influence the outcome of the model. Furthermore, the electoral system can have a significant influence on the dynamics of the model. Here, we only consider direct elections and leave other electoral systems, such as proportional elections \cite{PhysRevLett.99.138701}, for a later discussion. %\AS{\textit{Maybe here we can mention that a further study can consider the actual election system, see Ref. [sacling paper in proportional election]. This reference can be mentioned even in the first paragraph.}}

More specifically, we propose here generalizations to the SIRM model that are mathematically well-posed and conserve the idea behind the SIRM model. We also show that the SIRM model can be recovered as a limit of our model under specific assumptions. \JM{Hence, the SIRM model} turns out to be valid in the US presidential election case. \JM{The new feature of our model lies in an improved handling of the noise term. } Furthermore, we develop a calibration procedure required to actually apply our model to real voting data, opening the way to application of our model on various elections. We would like to underline the fact that VM-like models have fairly strong assumptions with respect to the behaviour of the voters. Some of these assumptions are unlikely to be true, but we nevertheless expect to be able to capture some statistical regularities in elections using this model. Such regularities have been examplified in \cite{PhysRevLett.99.138701,chatterjee2013universality,klimek2012statistical,Borghesi2010,borghesi2012election,enikolopov2013field,PhysRevE.60.1067,ARARIPE20094167,Bernardes2002,lyra2003generalized,PhysRevE.74.036112,araripe2006plurality,gonzalez2004opinion,andresen2008correlations,HERNANDEZSALDANA20092699,araujo2010tactical,KIM2003741}. 

\if 0
\begin{itemize}
\item Discuss what to do when we do not have actual data to work with
\item Full paper in itself to study at the topology implied by the commuting network (weighted directed graph with self-loop)
\item Dependence on initial voting conditions
\item We choose a particular topology and initial condition and study the influence of different parameter, aiming at providing a way to calibrate the model for applying it to real voting systems.
\end{itemize}
\fi

The remainder of this paper is organized as follows. In Section \ref{sec:GVMRM}, we present general requirements for the definition of a VM-like model based on a commuting network. In Section \ref{sec:SIRM}, we recall the SIRM model as presented in \cite{PhysRevLett.112.158701}. In Section \ref{sec:GSIRM}, we present our generalized version of the SIRM model and show that the SIRM model can be recovered as a limit of our model. In Section \ref{sec:NumExp}, we \JM{illustrate our model with numerical experiments on synthetic input data. We also illustrate a possible calibration procedure to set the parameters of our model.} A concluding discussion is provided in Section \ref{sec:Conclusion}.
%\begin{itemize}
%\item Is the voter model a model for voters?
%\item The Mallorca group answer: yes, with social mobility and noise
%\item Issues: The model breaks for small parties because of the noise term
%\item Are the results still valid for an arbitrary number of parties
%\end{itemize}
\section{Generalized Voter Model with Recurrent Mobility}\label{sec:GVMRM}
In this section, we discuss the general form that a VM-like model on the commuting network should have to be mathematically well-posed. To do so, we start by specifying what we understand as the commuting network and then introduce the relevant opinion space. Finally, we specify the dynamics that any \emph{Generalized Voter Model with Recurrent Mobility} (GVMRM) model should have.
%\subsection{Two parties formulation}
\subsection{Commuting network}
Consider a population of $N$ individuals divided into $M$ regions, that we call municipalities. This regions usually are electoral regions, such as counties, municipalities or states. Let $N_{ij}$ be the number of commuters between municipality $i$ and $j$, that is the number of individuals living in municipality $i$ and working in municipality $j$. The commuting network is constructed by taking the $M$ municipalities as nodes and the creating a weighted edges of weight $N_{ij}$ between municipalities $i$ and $j$. The commuting network is a directed weighted network with self-loops, since there are people who live and work in the same place. In the rest of the paper, we will refer to the sub-population $N_{ij}$ as the commuting cell $ij$.

From the $N_{ij}$ quantities, one can construct the number $N_i$ of people living in municipality $i$ by summing over the second index and the number $N'_{j}$ of people working in municipality $j$ by summing over the first index. We therefore defined
\begin{equation}
N_i := \sum_{j}N_{ij}
\quad \text{and} \quad
N'_j := \sum_{i} N_{ij}.
\end{equation}
The total population $N= \sum_{i}N_i = \sum_{j}N'_j$. The primed quantities refer to the working population.

\subsection{Opinion structure and vote shares}
Let us assume that the individual can choose between $K$ different opinions. We denote by $V_{ij}^k$ the number of people in commuting cell $ij$ that have opinion $k$. For consistency, we must have
\begin{equation}
\sum_{k=1}^K V_{ij}^k = N_{ij}.
\end{equation}
The state of the model is fully defined by the quantities $V_{ij}^k$. For convenience, we also introduce the vote shares
\begin{equation}
v_{ij}^k := \frac{V_{ij}^k }{N_{ij}}.
\end{equation}
With this definition one can easily show that
\begin{equation}
v^k_i = \sum_{j}\frac{N_{ij}}{N_i} v_{ij}^k
\quad
\text{and} 
\quad\textbf{}
{v'}^{k}_j = \sum_{i}\frac{N_{ij}}{N'_j} v_{ij}^k\label{eq:viLvjW}
\end{equation}
are the vote shares of opinion $k$ for the population living in municipality $i$ and for the population working in municipality $j$.

\if 0
\vspace{5cm}
For the moment, assume that there are only two possible opinions/parties $\vec 0$ and $\vec 1$. Let us define $V_{ij}$ as the number of people living in $i$ and working in $j$ holding opinion $\vec 1$. We also have that $N_{ij}-V_{ij}$ is the number of people living in $i$ and working in $j$ holding opinion $\vec 0$.

One can now define the vote shares $v_{ij}$ of opinion $\vec 1$ among people living in $i$ and working in $j$.

Similarly as before, we can define the corresponding quantities for people living in $i$ and people working in $j$
\begin{equation}
v_i = \sum_{j}\frac{N_{ij}}{N_i} v_{ij}
\quad
\text{and} 
\quad\textbf{}
v'_j = \sum_{i}\frac{N_{ij}}{N'_j} v_{ij}\label{eq:viLvjW}
\end{equation}
For $K$ options or parties, we introduce the quantity $v^k_{ij}$ to be the vote share of option $k$ in commuting cell $ij$. For consitency:
\begin{equation}
\sum_{k} v^k_{ij} = 1, \quad \forall i,j
\end{equation}
The quantities $v^k_{ij}$ are related to the number $V^k_{ij}$ of people holding opinion $k$ in commuting cell $ij$ by
\begin{equation}
v^k_{ij} := \frac{V^k_{ij}}{N_{ij}}.
\end{equation}
We additionaly introduce the vector quantity $\vec v_{ij}$ defined as
\begin{equation}
\vec v_{ij} = (v^1_{ij}, \dots,v^K_{ij})\quad \text{and}\quad \vec V_{ij} = (V^1_{ij}, \dots,V^K_{ij}).
\end{equation}
\fi

\subsection{Dynamics}
Let us now define the transition operators $R^{kk'}_{ij}$ 
\begin{equation}
R^{kk'}_{ij}= P[(V^k_{ij}, V^{k'}_{ij}) \to (V^{k}_{ij}-1, V^{k'}_{ij}+1)] = \frac{N_{ij}}{N}v^k_{ij}p^{k\to k'}_{ij},
\end{equation}
that defines the probability that an individual changes from opinion $k$ to opinion $k'$ in the commuting cell $ij$. For the left most term, the first factor is the probability to choose an individual in the commuting cell $ij$; the second factor is the probability that this individual holds opinion $k$ and the third factor is the probability that she changes opinion to $k'$.

%This quantity gives the probability that an individual holding opinion $k$ changes to $k'$ in commuting cell $ij$. The first factor is the probability to choose an individual in the commuting cell $ij$; the second factor is the probability that this agent holds opinion $k$ and the third factor the probability that she changes opinion to $k'$.

In general, any Voter Model variations defined on a commuting network can be defined by rates of the form
\begin{equation}\label{eq:generalRate}
R^{kk'}_{ij} = \frac{N_{ij}}{N}v^k_{ij}p^{k\to k'}_{ij},
\end{equation}
where $\vec p^{k \to }_{ij}=(p^{k\to 1}_{ij},\dots,p^{k\to K}_{ij})$ is the probability distribution to transition from any opinion $k$ to  other opinions in commuting cell $ij$.

In principle, any GVMRM model can be defined through Eq. \eqref{eq:generalRate} by specifying $p^{k\to k'}_{ij}$.

\section{The SIRM model}\label{sec:SIRM}
In this section, we reformulate the SIRM model in terms of (transition, raising and lowering) operators as done in \cite{sood2008voter} and recall the analysis that has been performed in \cite{PhysRevLett.112.158701}. 
\subsection{Formulation}
The SIRM model has been defined for 2 opinions. Let the 2 opinions be referred to as $\vec 0$ and $\vec 1$. The fundamental quantities of the models are given by
\begin{equation}
V_{ij} := V_{ij}^{\vec 1}\quad \text{and} \quad N_{ij}-V_{ij} := V_{ij}^{\vec 0}
\end{equation}
and the corresponding vote shares
\begin{equation}
v_{ij} := v_{ij}^{\vec 1}\quad \text{and} \quad 1-v_{ij} := v_{ij}^{\vec 0}.
\end{equation}
The vote share at home and at work can be computed through the expressions
\begin{equation}
{v}_i = \sum_{i}\frac{N_{ij}}{N_i} v_{ij}\quad\text{and}\quad v'_j = \sum_{i}\frac{N_{ij}}{N'_j} v_{ij}.
\end{equation}

In order to fully defined the SIRM model we need to specify the dynamics, that is the four operators $R^{kk'}_{ij}$, for $k, k' \in \{\vec 0, \vec 1\}$. Since the operators $R^{kk'}_{ij}$ gives the probability to change opinion from $k$ to $k'$, we must have
\begin{equation}
R^{\vec0,\vec 0}_{ij} = 1- R^{\vec0,\vec 1}_{ij} \quad \text{and}\quad R^{\vec1,\vec 1}_{ij} = 1- R^{\vec1,\vec 0}_{ij}.
\end{equation}
Therefore, it is sufficient to only specify $R^{\vec0,\vec 1}_{ij}$ and $R^{\vec1,\vec 0}_{ij}$ to fully define the model. In \cite{PhysRevLett.112.158701}, they called $R_{ij}:=R^{\vec0,\vec 1}_{ij}$ the raising operator and $L_{ij}:=R^{\vec1,\vec 0}_{ij}$ the lowering operator. With this notation, the SIRM model is defined by
\begin{equation}\label{eq:RL}
\begin{aligned}
R_{ij} 	&= \frac{N_{ij}}{N}\left\{(1-v_{ij})\left[\alpha v_i +(1-\alpha)v'_j\right] + \frac{D}{2}\eta^+_{ij}\right\},\\
L_{ij} &= \frac{N_{ij}}{N}\left\{v_{ij}\left[\alpha (1-v_i) +(1-\alpha)(1-v'_j)\right] + \frac{D}{2}\eta^-_{ij}\right\},
\end{aligned}
\end{equation}
where $\eta^{+,-}_{ij}$ are independent Gaussian white noises. The parameter $\alpha$ controls the ratio of interaction at home and at work and $D$ controls the magnitude of the noise.

By the definition of the raising and lowering operators, it is clear that for a well-posed model we need
\begin{equation}\label{eq:Cond1}
0\leq R_{ij}, L_{ij} \leq 1.
\end{equation}
For extreme values of $v_{ij}$ it is obvious that \eqref{eq:RL} does not satisfy \eqref{eq:Cond1}. For example, if $v_{ij}=0$ and $\eta^-_{ij}<0$, then $L_{ij} < 0$. We can conclude that the SIRM model is ill-posed and the issue originates in the noise term used. Remark that if $D=0$, then \eqref{eq:RL} is of the form \eqref{eq:generalRate} and the model is well-posed. 

\subsection{Analysis}\label{sec:Analysis}
%\subsubsection{Raising and Lowering operators and dynamics}
In this subsection, we consider the case of non-extreme vote shares as it has been developed for the US election case and recall the analysis performed in \cite{PhysRevLett.112.158701}. In this case, the problem with the noise mentioned before can be ignored. For clarity, we rewrite the raising and lowering operators as functions of $v_{ij}$, that is
\begin{equation}
R(v_{ij}):= R_{ij}\quad \text{and}\quad L(v_{ij}):= L_{ij}.
\end{equation}

Using these operators, we can write the corresponding balance equation or master equation
\begin{equation}
\begin{aligned}
p(v_{ij}, t+\delta t) = R(v_{ij}- \delta v_{ij})p(v_{ij}-\delta v_{ij},t) \\+ L(v_{ij}+ \delta v_{ij})p(v_{ij}+\delta v_{ij},t) \\+ [1-R(v_{ij}) - L(v_{ij})]p(v_{ij},t),
\end{aligned}
\end{equation}
where $\delta t = \frac1{N}$ and $\delta v_{ij} = \frac1{N_{ij}}$.

Expanding this equation to second order give a Fokker-Planck equation
\begin{equation}
\frac{\partial}{\partial t} p(v_{ij},t) = -\frac{\partial}{\partial v_{ij}}[\hat d(v_{ij})p(v_{ij},t)] + \frac{1}{2}\frac{\partial^2}{\partial v_{ij}^2} [\widehat D(v_{ij})p(v_{ij},t)],
\end{equation}
where the \emph{drift} coefficient, $\hat d(v_{ij})$, and the \emph{diffusion} coefficient, $\widehat D(v_{ij})$,  are given by 
\begin{equation}
\begin{aligned}
\hat d(v_{ij}) &= \frac{\delta v_{ij}}{\delta t}(R(v_{ij})-L(v_{ij}))\\
\widehat D(v_{ij}) &= \frac{\delta v_{ij}^2}{\delta t}(R(v_{ij})+L(v_{ij})).
\end{aligned}
\end{equation}

Equivalently, we can obtain a Stochastic Differential Equation (SDE)/ Langevin equation of the form
\begin{equation}\label{eq:SDE}
dv_{ij} = \hat d(v_{ij})dt + \sqrt{\widehat D(v_{ij})}dW^*_{ij}(t),
\end{equation}
where $dW_{ij}(t)$ is a white noise. There is a simple correspondence between $dW_{ij}$ and $\eta_{ij}$ given by
\begin{equation}
dW^{\circ}_{ij} = \eta^{\circ}_{ij}dt,
\end{equation}
where $\circ$ could be $*,',...$, see \cite[p.~90]{gardiner2009stochastic}.

For the raising and lowering operators of the SIRM model, we obtain
\begin{equation}\label{eq:dDSIRM}
\begin{aligned}
\hat d(v_{ij}) &= \alpha v_i + (1-\alpha)v'_j - v_{ij} + \frac{D}{\sqrt{2}} \eta_{ij}\\
\widehat D(v_{ij}) &= \frac{1}{N_{ij}}\left[(1-2v_{ij})[\alpha v_i + (1-\alpha)v'_j ]  + v_{ij} + \frac{D}{\sqrt{2}}\eta'_{ij}\right]
\end{aligned}
\end{equation}
where 
\begin{equation}\label{eq:sqrt}
\begin{aligned}
\frac{1}{\sqrt{2}} \eta_{ij} = \frac{1}{2}(\eta^+_{ij}-\eta^-_{ij})\\
\frac{1}{\sqrt{2}} \eta'_{ij} = \frac{1}{2}(\eta^+_{ij}+\eta^-_{ij})
\end{aligned}
\end{equation}
since Gaussian white noises add as standard deviations of normally distributed variables.

We would like to stress that both the drift coefficient, $\hat d(v_{ij}) $, and the diffusion coefficient, $\widehat D(v_{ij}) $, are stochastic functions because of the presence of $\eta$ terms. This is a consequence of working with stochastic rate of change. 
%\begin{itemize}
%\item Emphasize that $\hat d$ and $\hat D$ are stochastic functions
%\end{itemize}
\if 0
\subsection{Limitations}

One condition for this to be well-defined is: 
\begin{equation}
0\leq r^{k\to k'}_{ij} \leq 1, \forall k\neq k' \in \{\vec 0, \vec 1\}.
\end{equation}
As we have shown, this condition cannot be ensured with the original definition of the raising and lowering operators.

In the initial formulation the transition rate operators can be rewritten as
\begin{equation}
R^{kk'}(\vec v_{ij}) = \frac{N_{ij}}{N}\left[v^k_{ij}r^{k\to k'}_{ij} + \frac{D}{2}\eta^{kk'}_{ij}\right], \quad\forall k\neq k',
\end{equation}
where $\eta^{kk'}_{ij}$ are gaussian white noises. As discussed, this case is problematic since if $v^k_{ij} = 0$ the rate of change $R^{kk'}(\vec v_{ij})$ can either be positive, in which case there is a positive probability to decrease the number of people holding opinion $k$, which is already zero, leading to a negative number of people, or it can be negative, in which case the rate become negative.

There is therefore a need for new strategies to put noise on the SIRM model in a consistent way.
\fi
\section{The Generalized SIRM model}\label{sec:GSIRM}
In this section, we provide a generalization of the SIRM model (as a possible realization of the GVMRM model) that (i) can model any fixed number of opinions and (ii) has well-posed stochastic rates $R^{kk'}_{ij}$. We will show that the SIRM model can be recovered as a limit of our model when the vote share is far from the extreme cases (0 or 1).

\subsection{Generalization to $K$ opinions}
Let us start the discussion of the extension to $K$ opinions by considering deterministic rates. Under this assumption, we can define $r^{k\to k'}_{ij}$ to be the probability that an individual holding opinion $k$ will change to opinion $k'$ in the communting cell $ij$. The natural generalization to $K$ opinions is given by defining
\begin{equation}
r^{k\to k'}_{ij} := \alpha v^{k'}_i + (1-\alpha)v'^{k'}_j.
\end{equation}
The probabilities $r^{k\to k'}_{ij}$ are a specific instance of $p^{k\to k'}_{ij}$ in \eqref{eq:generalRate}. This probability does not depend on the current option $k$, since the copying process is unconditional. We keep the $k$ index, because the noise term that will be added will be $k$-dependent. For consistency, we have
\begin{equation}
\sum_{k'}r^{k\to k'}_{ij} = 1, \quad \forall i,j,k,
\end{equation}
since $r^{k\to k'}_{ij}$ forms a probability distribution.

\subsection{Noise handling}
We now discuss the different possible options to add noise on the rates $r^{k\to k'}_{ij}$, while keeping the probability distribution property. We discuss various ways to modify the rates both in a deterministic way and in a stochastic way to construct general noisy rates. The way noise has been added to the VM mainly relies on a deterministic modification of the rates \cite{scheucher1988soluble,granovsky1995noisy}. 

\subsubsection{Adding free will}
One possibility to add a kind of noise is to introduce ``free will''. Free will can be encoded in a fixed probability to randomly change state. Let us assume that there are $K$ options, then one can choose the uniform probability to change opinion.% as $f^{k\to k'}_{ij} = \frac{1}{K}$. 

One can add free will to the SIRM model without noise by redefining $R^{kk'}(\vec v_{ij})$ as
\begin{equation}
R^{kk'}(\vec v_{ij}) =  \frac{N_{ij}}{N}v^k_{ij}\left[(1-\gamma)r^{k\to k'}_{ij} + \gamma \frac{1}{K} \right],
\end{equation}
where $\beta$ controls the strength of the free will term. In the general notation introduced in \eqref{eq:generalRate}, we have
\begin{equation}\label{eq:freewill}
p_{ij}^{k\to k'} =(1-\gamma)r^{k\to k'}_{ij} + \gamma \frac{1}{K}.
\end{equation}

This modification is interesting, because it allows for opinions that went extinct to reappear in the system. This can also be used to simulate the creation of a new party. Without such a term, the number of party is strictly decreasing. In order to simulate real life elections, such a noise (not necessarily uniform) would be needed.

\subsubsection{Adding intra-commuting cell influence}
In the context of GVMRM models, the commuting pattern is important. One could assume that agents living in the same commuting cell interact more often than people who do not. According to \cite{feld1981focused,feld2009homophily} people sharing more loci (place of living, place of work, etc.) have more influence on one another. We can take this effect into account by adding more weight to intra-commuting cell neighbors. Similarly to what has been done for free will, we can introduce an additional probability to change from opinion $k$ to $k'$ in commuting cell $ij$ proportional to $v^{k'}_{ij}$. The model can now be modified through
\begin{equation}\label{eq:intra}
R^{kk'}(\vec v_{ij}) =  \frac{N_{ij}}{N}v^k_{ij}\left[(1-\beta)r^{k\to k'}_{ij} + \beta v^{k'}_{ij}\right],
\end{equation}
where $\beta$ controls the strength of the local interaction term.

%Reference to OH of analytical sociology loci theory p.85 see also Feld 1981,  Feld and Grofman 2009. \cite{feld1981focused,feld2009homophily}

%To add intra-commuting cell influence, one can, instead of a uniform free will, assume that people mainly interact with the people in their own commuting cell. Let us define a probability $g^{k\to k'}_{ij}= v^{k'}_{ij}$. The model can now be modified through
%\begin{equation}
%R^{kk'}(\vec v_{ij}) =  \frac{N_{ij}}{N}v^k_{ij}\left[(1-\gamma)r^{k\to k'}_{ij} + \gamma v^{k'}_{ij}\right],
%\end{equation}
%where the $\gamma$ parameter controls the strength of the local interaction term.

In principle, any noise originating in a probability distribution over different opinions can be added in this way. Note that most of the time, the form of the probability distribution does not depend on the initial opinion $k$ but only on the final opinion $k'$ and might even be fully independent of opinions as in the case of free will. Furthermore, all these modifications are deterministic in the sense that they the transition rates and deterministic. This is a major difference between the original SIRM model and our formulation. The next step is to obtain in a consitent way stochastic versions of the rates $R^{kk'}(\vec v_{ij})$.

\subsubsection{Putting noise on a probability distribution: the Dirichlet solution}
In order to construct stochastic rates in a consistent manner, we need to find a way to transform a probability distribution into another that is stochastically perturbed from the initial one. To do so, we rely on probability distributions over the simplex. We have a few available choices, such as the Dirichlet distribution, the multinomial distribution or the lognormal distribution. The  multinomial distribution has been chosen in \cite{PhysRevE.95.022308} in the context of language evolution to model finite length conversation and it turns out that the structure of the covariance matrix of the multinomial distribution is proportional to that of the Dirichlet distribution.

We choose to use the Dirichlet distribution, because (i) it is a continuous distribution (not discrete like the multinomial); (ii) the correspondence between the initial probability distribution and the parameter of the Dirichlet distribution is easy to define; (iii) it has well-defined moments that can be used in the analysis and (iv) it has the aggregation property:
\begin{equation}
\begin{aligned}
&\text{If }(X_1,\dots,X_K)\sim \mathcal D(\alpha_1,\dots,\alpha_K) \text{ then }\\ &(X_1,\dots,X_i+X_j,\dots,X_K)\sim \mathcal D(\alpha_1,\dots,\alpha_i+\alpha_j,\dots,\alpha_K).
\end{aligned}
\end{equation}
This means that the model is compatible with aggregation of opinions. Aggregation of parties leads to a consistent formulation, which would not be true when using other distributions, such as the lognormal distribution. 
%which means that if we reduce the dimension of a stochastic vector sampled from this distribution, it still follows a Dirichlet distribution, whose parameter can be obtained by summing the corresponding parameters. 
These four properties ensures that the resulting model is consistent, well-posed and robust with respect to opinion aggregation.

%The main question addressed in this section is: How to put noise on a probability distribution, under the constraint that the resulting perturbed quantities are still a probability distribution?

%Our answer of this question relies on the Dirichlet distribution. 
Let us assume that a sample $\vec X$ from the Dirichlet distribution of parameter $\boldsymbol \alpha$ is given by $\vec X \sim {\mathcal D}(\boldsymbol \alpha)$. If we define $\bar \alpha= \sum_{k} \alpha_k$ then we can write the expected value 
\begin{equation}
\mathbb E(X_k) = \frac{\alpha_k}{\bar\alpha}
\end{equation} 
and the covariance matrix
\begin{equation}
\vec C = \frac1{\bar\alpha^2(1+\bar\alpha)}({\rm diag}(\boldsymbol \alpha) - \boldsymbol \alpha\boldsymbol \alpha^T) . 
\end{equation}
%\quad \text{and}\quad {\rm Var}(X_k) = \frac{\alpha_k(\bar\alpha-\alpha_k)}{\bar\alpha^2(1+\bar\alpha)}, {\rm Cov}(X_k,X_{k'}) = \frac{-\alpha_k\alpha_{k'}}{\bar\alpha^2(1+\bar\alpha)}

Let us now assume that we want to put noise on a probability distribution $\vec p = (p_1,\dots,p_K)$ with the consistency relation $\sum_k p_k = 1$ and $p_k\geq 0,\ \forall k$. Then we can define a noisy probability distribution $\widetilde{\vec p}_{\tilde D} \sim {\mathcal D}(\vec p/{\tilde D})$, where $\mathcal D$ stands for the Dirichlet distribution. It is straightforward to verify that 
\begin{equation}
\mathbb E(\widetilde{ p}_{{\tilde D},k}) = p_k\quad \text{and}\quad \vec C = \frac{{\tilde D}}{{\tilde D}+1}({\rm diag}(\vec p) - \vec p\vec p^T).
\end{equation}

When ${\tilde D}$ is small the variance is small. When ${\tilde D} \to \infty$, then the prefactor tends to $1$, which is the largest possible variance on the simplex.

This procedure to get a stochastic version of a probability distribution can be used to add stochasticity to the rates of the GVMRM. In the rest of this paper, we consider the following model:
\begin{equation}\label{eq:extSIRM}
p^{k\to k'}_{ij} = (1-\beta-\gamma)r^{k\to k'}_{ij} + \beta \widetilde v^{k'}_{ij,{\tilde D}} + \gamma/K,
\end{equation}
where
\begin{equation}
\widetilde{\vec v}_{ij,{\tilde D}} = {\mathcal D}(\vec v_{ij}/{\tilde D}).
\end{equation}
We also require $\beta + \gamma \leq 1$ and $\beta,\gamma \geq 0$. The parameter $\beta$ represents the stength of the intra-commuting cell influence as defined in Eq.~\eqref{eq:intra}, the parameter $\gamma$ controls the intensity of free will as defined in Eq.~\eqref{eq:freewill} and the parameter $\tilde D$ controls the level of stochasticity of the rate.

We stress that the model defined through \eqref{eq:extSIRM} includes both a free will term and an intra-commuting cell term. Furthermore, we add some stochasticity to the intra-commuting cell term to introduce stochastic rates. The idea to only add noise in the intra-commuting cell term is motivated by the fact that we want a local noise term, that is a noise that is indepedent between different commuting cell as is the case in \cite{PhysRevLett.112.158701}. Other choices could have been made and we will not investigate them in this paper.

\if 0

\subsection{Analysis}
%\vspace{5cm}
\begin{itemize}
\item Write the master equation associated with our model
\item Obtain the associated SDE, under Normal approximation of the Dirichlet distribution.
\item Discuss shortly the dynamics implied
\end{itemize}

\fi
\subsection{The SIRM model as a limit}\label{ssec:SIRMlimit}
In the case of two variants with the transition probability defined in \eqref{eq:extSIRM}, we can apply the same procedure in Section \ref{sec:Analysis} and compute the drift and diffusion coefficients, $\hat d(v_{ij})$ and $\widehat D(v_{ij})$, respectively. We obtain the following terms:
\begin{equation}\label{eq:dDSIRM_new}
\begin{aligned}
\hat d(v_{ij}) &=(1-\beta-\gamma)[r^{\vec 0 \to \vec 1}_{ij}  - v_{ij}] \\ 
	&\quad + \beta(\widetilde v_{ij,\tilde D}-v_{ij}) \\
	&\quad+ \gamma\left(\frac12-v_{ij}\right)\\
\widehat D(v_{ij}) &=\frac{1}{N_{ij}}\Big[(1-2v_{ij})\left[(1-\beta-\gamma)r^{\vec 0 \to \vec 1}_{ij} + \beta \widetilde v_{ij,\tilde D}\right]\\&\quad +(1-\gamma)v_{ij} + \frac{\gamma}{2}\Big],
\end{aligned}
\end{equation}
where 
\begin{equation}
r^{\vec 0 \to \vec 1}_{ij} = \alpha v_i + (1-\alpha)v'_j.
\end{equation}
 Equations \eqref{eq:dDSIRM_new} are generalizations of Equations \eqref{eq:dDSIRM}.

It is important to note that the term $\widetilde v_{ij,\tilde D}$ is a random variable, since it is sampled from a Dirichlet (in this case Beta) distribution. In the following discussion, we will only consider the $\hat d(v_{ij}) $ term because the diffusion term, $\widehat D$, in the original version of the SIRM model was neglected, since it is proportional to $\frac{1}{N_{ij}}$ (and $N_{ij}$ is usually large). To get our formulation in a similar form as \eqref{eq:dDSIRM}, we can use the normal approximation of the Dirichlet distribution and approximate 
\begin{equation}
\widetilde v_{ij,\tilde D} \approx v_{ij} + \sqrt{\frac{\tilde D}{\tilde D+1}}\sqrt{v_{ij}(1-v_{ij})}\eta_{ij},
\end{equation}
where $\eta_{ij}$ is a delta correlated white noise. Under this approximation, the second term in Eq.~\eqref{eq:dDSIRM_new} turns into a pure noise term and we have
\begin{equation}
\begin{aligned}
\hat d(v_{ij}) &= (1-\beta-\gamma)[r^{\vec 0 \to \vec 1}_{ij}  - v_{ij}] \\
&\quad+ \beta\sqrt{\frac{\tilde D}{\tilde D+1}}\sqrt{v_{ij}(1-v_{ij})}\eta_{ij} \\
&\quad+ \gamma\left(\frac12-v_{ij}\right).
\end{aligned}
\end{equation}
If we now assume that $\gamma = 0$ and  $v_{ij} \approx \frac12$ so that the square root $\sqrt{v_{ij}(1-v_{ij})} \approx \frac{1}{2}$, we get
\begin{equation}
\hat d(v_{ij}) = (1-\beta)[r^{\vec 0\to \vec 1}_{ij}  - v_{ij}] + \frac{\beta}{2}\sqrt{\frac{\tilde D}{\tilde D+1}}\eta_{ij}.
\end{equation}
If we then set $\beta = \frac12$ we get
\begin{equation}
2\hat d(v_{ij}) = [r^{\vec 0\to \vec 1}_{ij}  - v_{ij}] + \frac{1}{2}\sqrt{\frac{\tilde D}{\tilde D+1}}\eta_{ij},
\end{equation}
which, up to factor $2$ that can be eliminated by a rescale of time, we have a very similar expression as Eq.~\eqref{eq:dDSIRM}. Furthermore, we have the approximate correspondance:
\begin{equation}\label{eq:DtildeD}
D \approx \frac{1}{\sqrt2}\sqrt{\frac{\tilde D}{1+\tilde D}}\quad\text{ and }\quad\tilde D \approx \frac{2D^2}{1-2D^2}.
\end{equation}

With the adequate correspondance of the noise factor, the parameter $\beta$ only acts as a time scale and since the time correspondence between the model and the real data has to be calibrated for, its influence does not impact the general result  (unless $\beta = 1$ in which case the time scale factor become infinite).

\section{Numerical results}\label{sec:NumExp}
In this section, we study the behaviour of the generalized SIRM model defined by Eq. \eqref{eq:extSIRM}. As stated in Section \ref{sec:Intro}, we do not apply the model to real data, but to generated data. From now on, we set $\alpha$ and $\beta$ to $\frac12$ for simplicicty. In \cite{PhysRevLett.112.158701}, the actual value of $\alpha$ has been shown not to have a strong influence on the dynamics and $\beta$ mainly acts as a time scale, as discussed in Section~\ref{ssec:SIRMlimit}. Thus, it is sufficient to study the influence of the two other parameters added to the model: $\tilde D$ and $\gamma$. In order to run the model, we need to specify initial conditions, that is, we need to specify the commuting network and the initial vote share distribution. When applying the model to real election data, the initial conditions are inferred from the data. The focus of our numerical experiments is to study the influence of the two types of noise (free will and stochastic rates) added to the recurrent mobility component. 

The commuting network we choose for numerical experiments is based on a fully connected directed graph with self-loops. The nodes represents living/working places and we associate with every link of that network a number of commuters $N_{ij}$, randomly chosen between $0$ and $100$. We then multiplied the $N_{ii}$ quantities by a factor $100$ to account for the fact that the majority of people live and work in the same place. The simulated commuting network has 25 nodes and a total population of $N=160 286$.

The initial voting distribution is constructed by partitioning the commuting populations $N_{ij}$ into the number $K$ of possible options (chosen to be $K=10$ for this paper). The choice of $10$ parties is not arbitrary but motivated by the fact that in many contries, the number of parties with representation in the government is roughly $10$. For example, Iceland has $8$ elected parties and Sweden has $9$. Even in the US, where two parties dominates, if we look at the parties with represented at the state level, the number of parties is also $8$. Furthermore, this number of parties does not seem to depend on the size of the population, as demonstrated by the chosen examples. \JM{For instance, populations range from the order of $10^5$ for Iceland to the order of $10^8$ for the US, covering 3 orders of magnitude.} The partitioning is done by sequentially partitioning the $N_{ij}$ into the different options $V^k_{ij}$: $V^1_{ij}$ is uniformily sampled between $0$ and $N_{ij}$, then the number of partisans of the second party $V^2_{ij}$ is uniformily sampled from the remaining population $N_{ij}-V^1_{ij}$. This procedure is repeated until all parties but one have been assigned and the remaining population is assigned to the last party. This procedure generates a party distribution characterized by 1 or 2 large parties and many small parties, similar to actual party distributions \cite{hart1983endogenous}.

%Here I need to find out what distribution I actually used to generate the initial vote shares, I know the algorithm, but I need to find out the actual distribution
Since we are dealing with many parties, we need to use an adapted visualisation technique. We choose to plot the standard deviation $\sigma$ of a party vote share as a function of its mean $\mu$. This type of representation provides a good picture of the distribution of parties. In many previous studies \cite{Borghesi2010,borghesi2012election,PhysRevLett.112.158701}, it has been observed that the distribution of party size is roughly stationary over multiple elections. This fact will be used to calibrate our model.

In the rest of this section, we discuss the influence of $\tilde D$ when $\gamma=0$, which is the closest case to the SIRM model. This leads to a calibration procedue for $\tilde D$. We then investigate the infuence of free will on the dynamics of the system. See Supplemental Material at [URL will be inserted by publisher] for the implementation of our model in Python and details about the computations, the code is also available in \footnote{The code can be downloaded from \href{https://www.researchgate.net/publication/322721295_Supplemental_material_for_Social_Influence_with_Recurrent_Mobility_with_multiple_options
}{here.}}.

%https://www.researchgate.net/publication/322721295_Supplemental_material_for_Social_Influence_with_Recurrent_Mobility_with_multiple_options

%A paragraph on the visualization of the distribution of parties (standard deviation as a function of mean)

\if 0
Plan of the section: Start with $\gamma = 0$...
\begin{itemize}
\item Discuss the conditions without data, what to choose...
\item We can play both with the topology and the initial vote share
\item Describe the commuting network used in the simulations
\item Describe the initial condition for the vote shares
\item Describe the visualization technique with many parties (variance of parties as a function of the mean)
\item Plan of the rest of the section
\end{itemize}
\fi

\subsection{Influence of $\tilde D$ ($\gamma = 0$)}
\begin{figure*}
\includegraphics[scale=0.5]{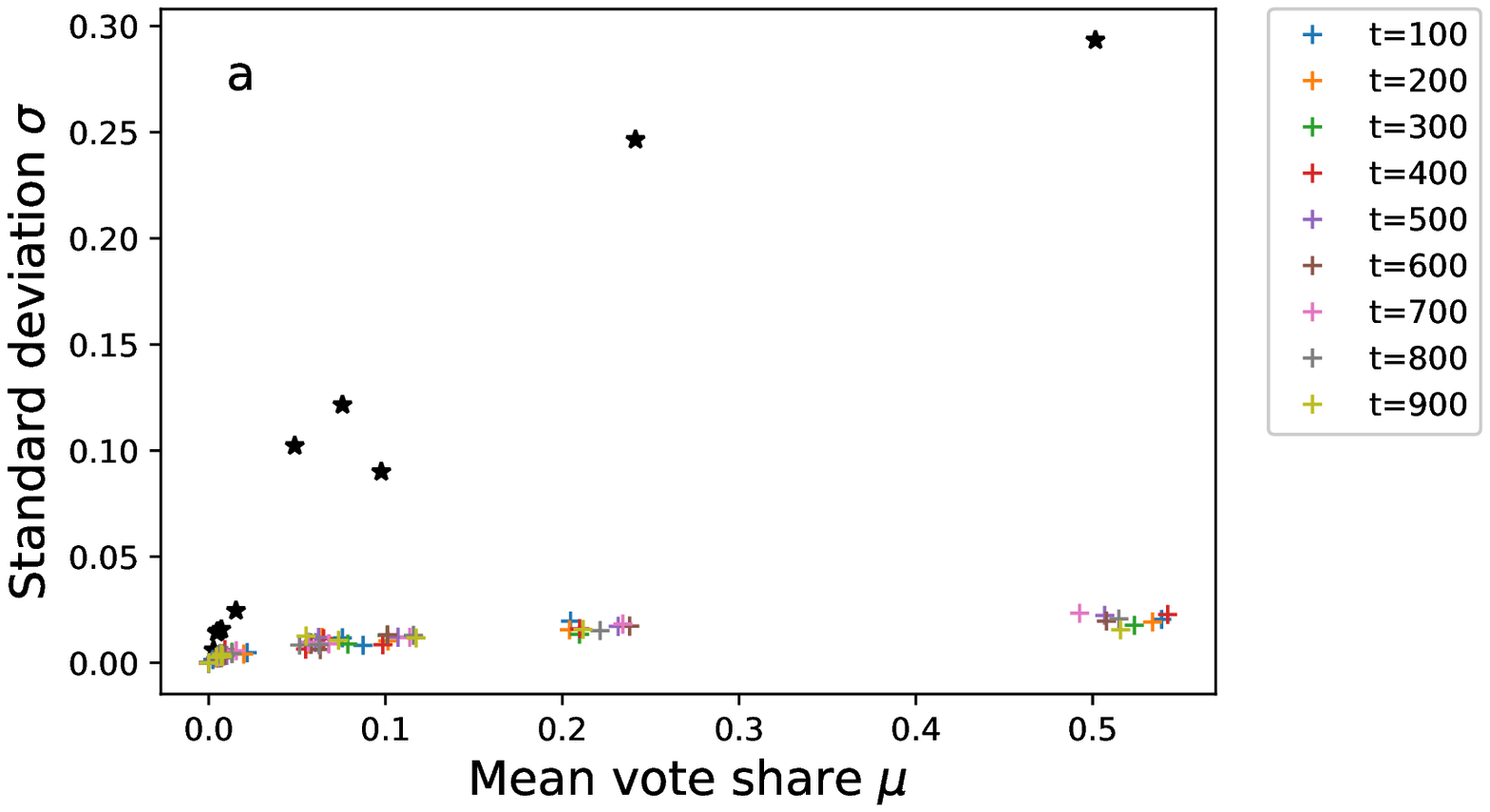}
\includegraphics[scale=0.5]{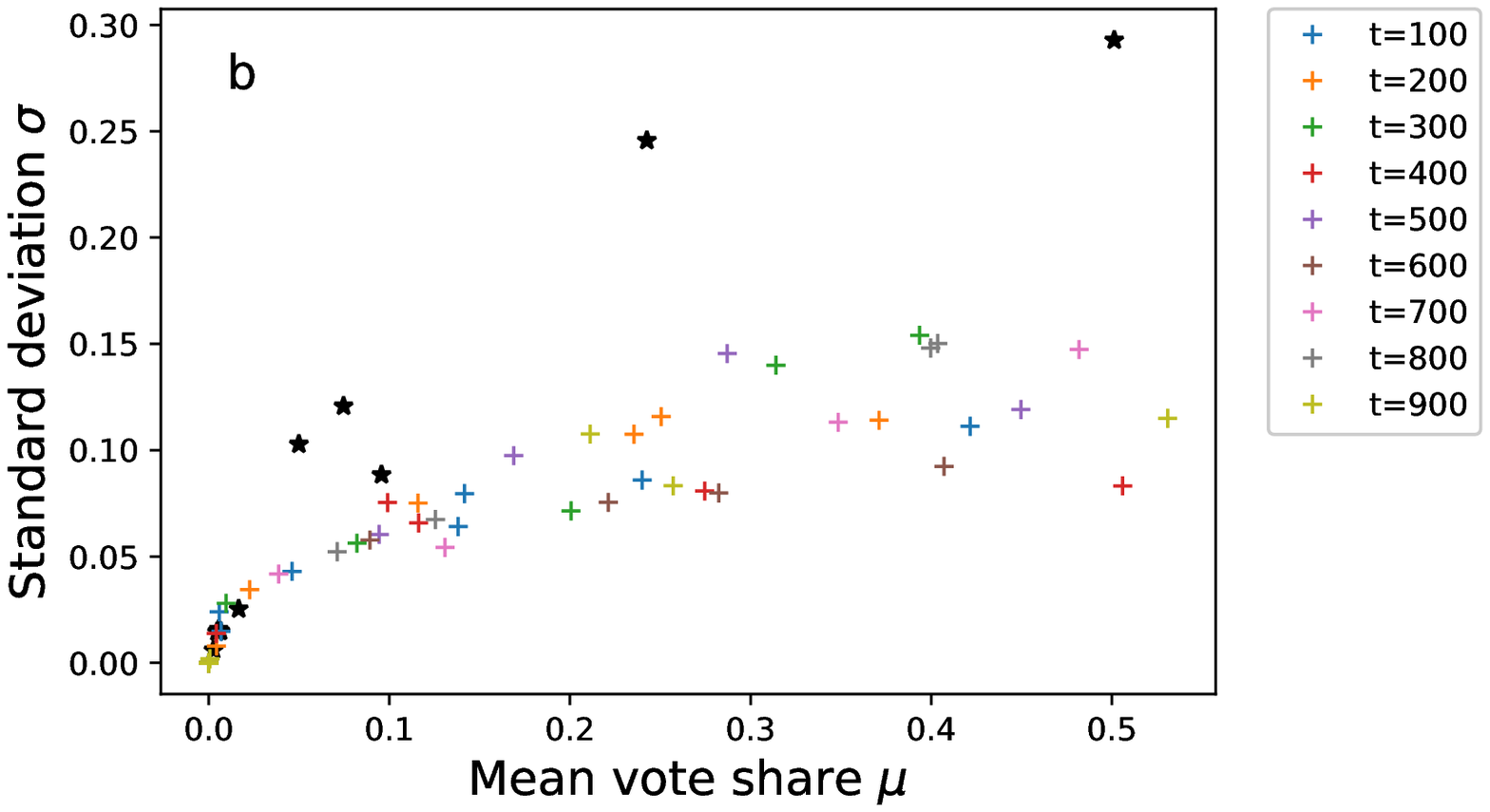}

\includegraphics[scale=0.5]{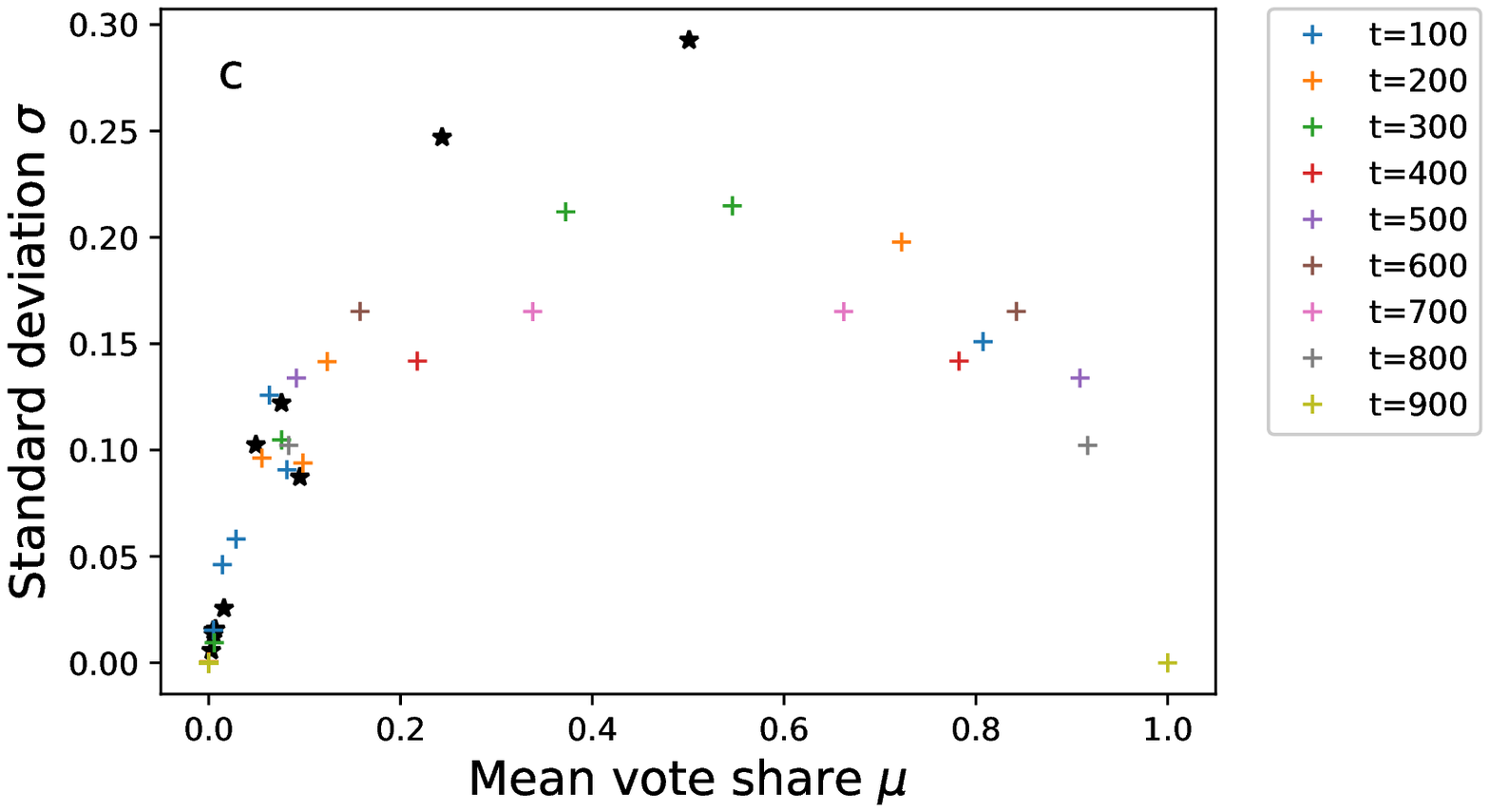}
\includegraphics[scale=0.5]{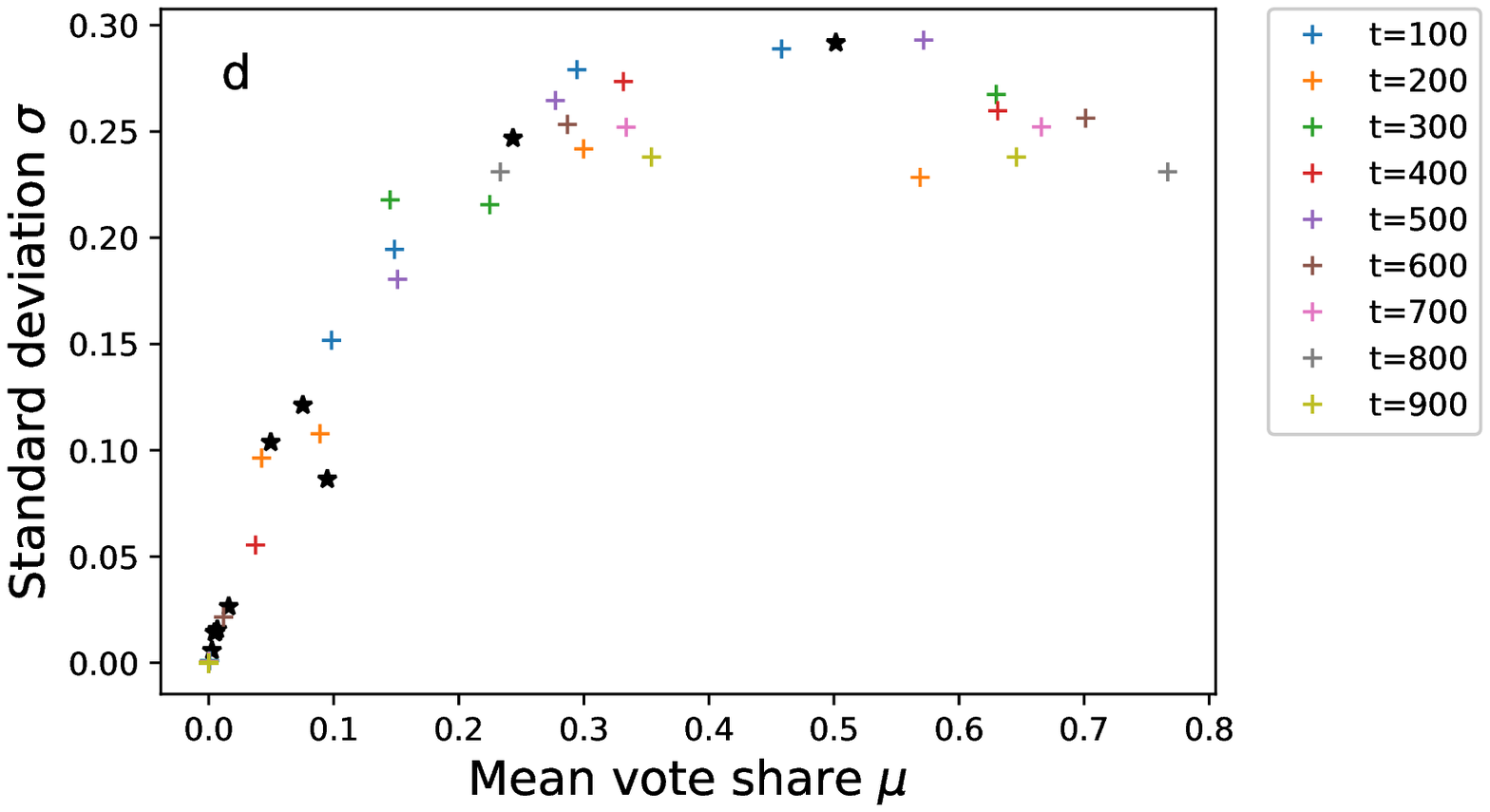}
\caption{Dependence of the results as a function of $\tilde D$. We display the results for $\tilde D =0, 0.1, 0.3$ and $1.0$ in panels {\bf a}, {\bf b}, {\bf c} and {\bf d}, respectively.}
\label{fig:tildeD}
\end{figure*}

Let us start by simulating the system for various values of $\tilde D$. Results are displayed in Fig.~\ref{fig:tildeD}. The initial party distribution is displayed by stars and party distribution at later times by plusses. We observe that for small values of $\tilde D$, the standard deviation is quickly reduced, indicating a convergence over the population to the given averaged values of the party vote shares. When increasing the value of $\tilde D$, the standard deviation increases as well, indicating more heterogeneity in the system. This is expected, since $\tilde D$ controls the importance of the noise and more noise leads to larger standard deviations. 

\begin{figure}
\includegraphics[width=\columnwidth]{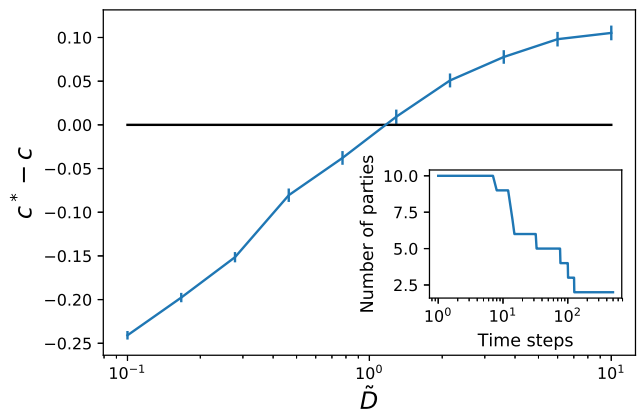}
\caption{Calibration curve for $\tilde D^*$. Inset: Decrease of the number of parties computed for the critical value $\tilde D^*$.}
\label{fig:Dstar}
\end{figure}

At some critical value $\tilde D^*$ of the diffusion coefficient, the evolution predicts a fairly stationary dynamics and the standard deviations stays at the same level as that of the initial conditions. The stationnarity criterion can be used to calibrate the model. For instance, we can fit the initial data to a specific function and repeat the fit at later times. If we use a function with a single parameter, we can run the model for different values of $\tilde D$ and choose the specific value $\tilde D^*$ when the coefficients of the fit match. We assume that the functional form of the standard deviation dependence on the mean follows a function $f(x) = c\sqrt{x(1-x)}$. In this case, the only parameter of this function is $c$. In Fig.~\ref{fig:Dstar}, we report the calibration curve for $\tilde D$, which leads to a critical value of $\tilde D^* \approx 1.12$.

The critical value $\tilde D^*$ can be related to the value of $D$ used in the original SIRM model \cite{PhysRevLett.112.158701} using Eq. \eqref{eq:DtildeD}. In our case, we obtain a critical value of $D^* \approx 0.51$, which is one order of magnitude larger than the value of $D^*_{\rm SIRM} = 0.03$ obtained in \cite{PhysRevLett.112.158701}. This indicates that the random assignment of vote shares used in our numerical experiments leads to a larger noise value than actual vote share distribution. What this result means is that in order to conserve the stationarity of random initial data, then the noise level should be significantly higher than to conserve real initial conditions.

One of the drawbacks of using a model without a free will term ($\gamma = 0$) is that the number of parties decreases during the evolution of the system, see the inset of Fig.~\ref{fig:Dstar}, where the decrease of the number of parties is displayed ($\tilde D = \tilde D^*$). As can be seen, many parties go extinct fairly quickly and the actual number of remaining parties turns out to be quite small (2 in this case). %In the next section, we investigate in detail the influence of the free will term on the dynamics.

\if 0
\begin{itemize}
\item This is the setting closest to the SIRM paper.
\item By varying $\tilde D$, the variance increases with $\tilde D$ and there is a level at which the stationarity condition is achieved. (graphs similar to the Orebro talk)
\item Stationarity condition can be used to calibrate the model: Plot the calibration curve for the chosen network and initial condition of vote shares.
\item But number of parties is decreasing with time (graph of the number of parties as a function of time)
\end{itemize}
\fi
\subsection{Influence of free will}
\begin{figure}
\includegraphics[width=\columnwidth]{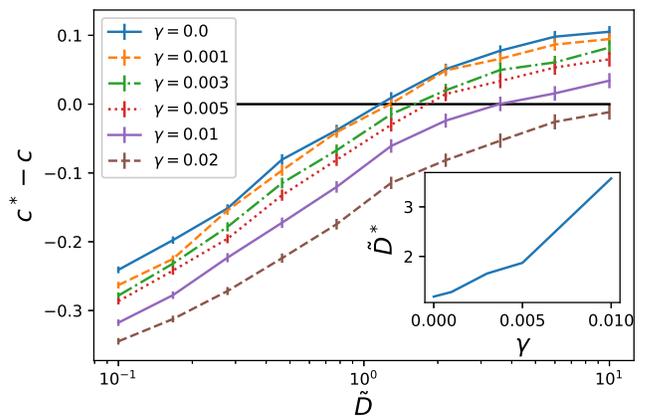}
\caption{Effect of $\gamma$ on the calibration curves of $\tilde D^*$. Inset: functional dependance of $\tilde D^*$ on $\gamma$.}
\label{fig:GammaCalibration}
\end{figure}

For the study of the influence of free will, we consider the following settings. We set
 $\alpha = 0.5-\gamma/2$, $\beta = 0.5-\gamma/2$ and modify $\tilde D$ for different values of $\gamma$. The presence of free will allows extinct parties to re-enter the dynamics. Results of calibration curves are displayed in Fig.~\ref{fig:GammaCalibration}. We observe that the critical value of $\tilde D^*(\gamma)$ increases as a function a $\gamma$, see the inset in Fig.~\ref{fig:GammaCalibration}, and when $\gamma$ exceeds a thereshold, the calibration procedure fails (no intersection between the curves). This is due to free will dominating the dynamics and leading to a well-mixed state. 

\begin{figure}
\includegraphics[width=\columnwidth]{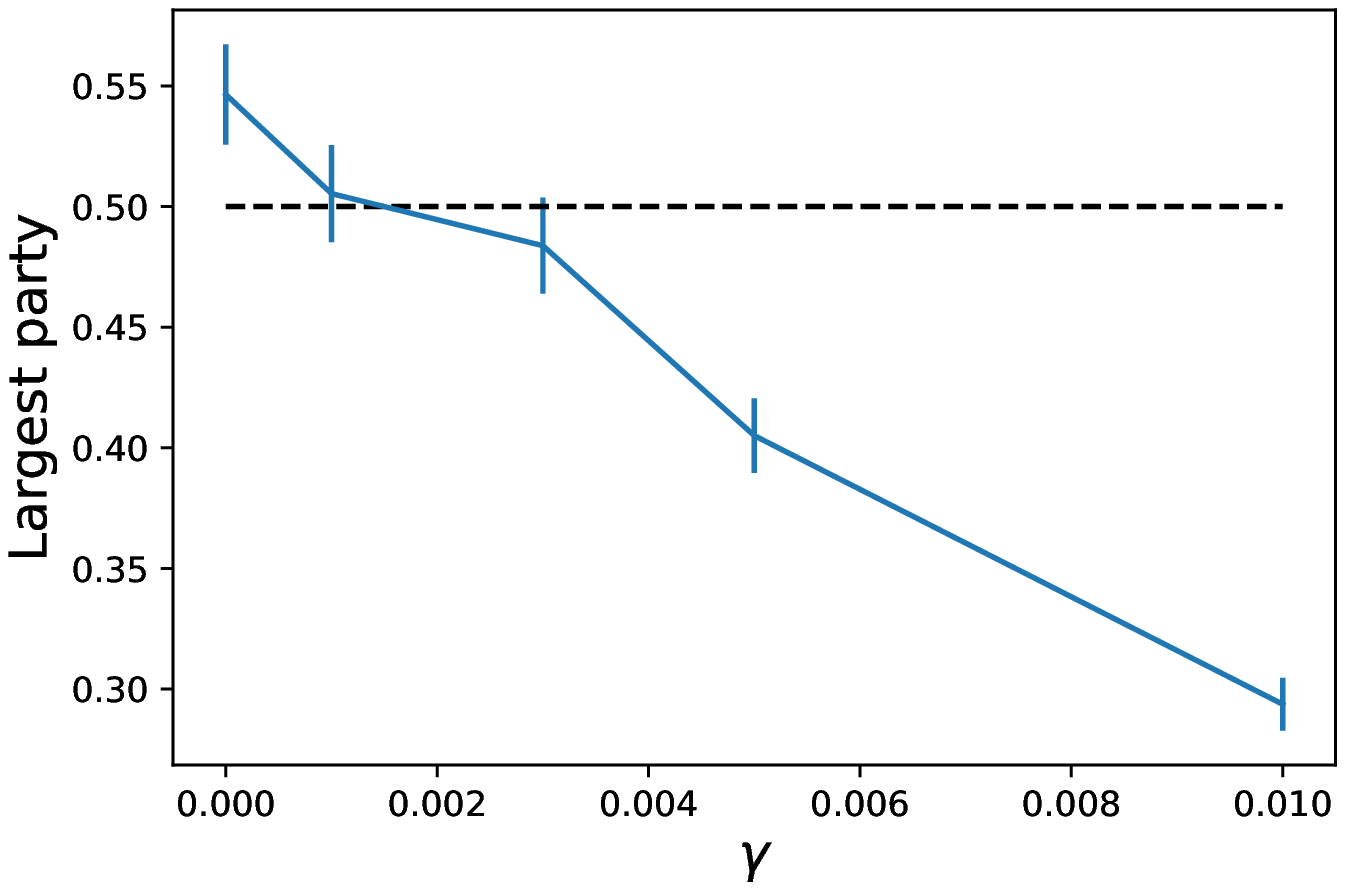}
\caption{Calibration curve of $\gamma^*$ based on the stationarity of the largest party size.}
\label{fig:largestParty}
\end{figure}

In order to fully calibrate the model, the intensity of free will should be specified. In order to fix the free will intensity, we can use the relationship between $\tilde D^*(\gamma)$ and $\gamma$ and run the model for various values of $\gamma$. The criterion used to choose the optimal value $\gamma^*$ for the free will component is obtained by looking at the maximum expected size of the largest party. In Fig.~\ref{fig:largestParty}, we display the calibration curve for $\gamma$. To obtain this curve, we run the model 50 times for 100 time steps for each values of $\gamma$ and averaged the largest party sizes. The error bars display the standard deviation of the largest party sizes. We observe that the maximum party size is a decreasing function of $\gamma$, which justifies its use as a calibration criterion, but the standard deviation for low values of $\gamma$ is rather large and leads to a big uncertainity on the value of $\gamma^* \approx 1.5\cdot 10^{-3}$. This issue can be solved by either averaging more realizations of the process or by running it for longer periods and, ideally, by doing both. For this paper, there is no need for precision, since we are interested in illustrating the behaviour of the model and proposing a calibration procedure. More accuracy could be needed for application to real voting data.

\if 0
\begin{itemize}
\item Presence of free will permits the extinct parties to reappear in the simulation
\item Influence of $\gamma$ on the calibration curve: 
\begin{itemize}
\item Effective drift attracting the parties toward each other, leading to a full mixing of opinions in the population
\item Shifting the critical value for $\tilde D$ to larger values. 
\end{itemize}
\item There is a thereshold above which the calibration procedure fails, because free will is too strong.
\item Test whether the vote share of the largest opinion is monotonically decreasing with $\gamma$
\item That would provide a criteria to choose the intensity of the free will term and fully define the model.
\item Influence of topology, initial voting shares, and other parameters ($\alpha, \beta$).
\end{itemize}
\fi

In this section, we have provided a calibration procedure for $\tilde D$ and $\gamma$, we did not explore in details the effect of $\alpha$ and $\beta$. This is left for another study. According to \cite{PhysRevLett.112.158701}, the effect of $\alpha$ on the dynamics is negligible and $\beta$ mainly acts as a time scale, see Sec.~\ref{ssec:SIRMlimit}. Using this calibration procedure, this model can be applied to real data, providing that the initial vote share distribution and the communting network are known.

\section{Conclusion}\label{sec:Conclusion}
In this paper, we have provided a generalized formulation for the Voter Model with Recurrent Mobility and illustrated the issues with its definition in \cite{PhysRevLett.112.158701}. We then developed a general procedure to add two types of noise ($\tilde D$ and $\gamma$) in this model and provided a generalization of the SIRM model. We stress the importance of using two different types of noise in the system: the free will term ($\gamma$) and stochastic rate of opinion change ($\tilde D$).

We have illustrated how the SIRM model could be recovered as a limit of our generalized model and examplified a possible calibration procedure for $\tilde D$ and $\gamma$ through numerical experiments. Note that the diffusion constant ($\tilde D^*$), calculated when $\gamma=0$, from our model is more than ten times larger than the corresponding value ($D^*_{\rm SIRM}$) obtained for the SIRM model \cite{PhysRevLett.112.158701}. This indicates that a calibration procedure based on stationarity requires more noise for random initial conditions than for real data, which is expected. Updating the SIRM model and adapting it to many options is a required step for the application of this model to real elections with many parties, such as the Swedish electoral system, but see \footnote{In real election systems, like proportional elections, the voting dynamics is more complex and further work should be done to adapt our model to that case, see, for example, \cite{PhysRevLett.99.138701}.}. The model is now ready to be applied to actual data and contribute to the question ``Is the Voter Model of Model for Voters?'' for situations when the vote share for a party is extreme (0 or 1).% \JM{In real election systems, like proportional elections, the voting dynamics is more complex and further work should be done to adapt our model to that case, see, for example, \cite{PhysRevLett.99.138701}.}

Free will is crucial to get a dynamics that can be related to real opinion dynamics, since new opinions do arise. However, the assumption that transition rates between different opinions are uniform is likely to be too strong, see \footnote{Opinions and parties are not fully independent of one another and opinion structures should be introduced at some point. In any VM-like model the opinion structure is ignored and opinion change only depends on demographics.}. One natural extension of this work would be to investigate the influence of structure in the opinion space by introducing constrains on the interactions, as done in \cite{vazquez2003constrained} or by adapting the bounded confidence model of Hegselmann and Krause \cite{HegselmannKrause02} to the Voter Model. One can also imagine applying this type of model in situations where opinions are represented in higher dimensional opinion spaces. The Axelrod model of cultural evolution \cite{doi:10.1177/0022002797041002001} is an example of how this could be done.

This paper hints to many other possible developments of this work, since the rates can be modified in many ways. One could, for example, use the Dirichlet distribution to add noise on other components of the rate and include time dependent or spatially dependent rates to account for varying and heterogeneous socio-economic factors that might have an influence on the dynamics of the system. Furthermore, the influence of the commuting network can be studied through numerical experiments. The network used in this paper is just one among many possibilities and it would be interesting to understand the influence of the topology of this weighted directed graph on the dynamics of the system. %\AS{\textit{We may mention how to later consider real election system, like proportional system... Could we mention that we are really planning to write a paper with Stockholm data very soon?}}
%\JM{\it I think that PRE does not allow to mention papers that are going to be submitted...}

\section*{Acknowledgements}
Earlier versions of this work have been presented at seminars at the University of Uppsala and at the University of \"Orebro. We would like to thank the audience of these seminars for interesting comments and discussions. In particular, we would like to thank Prof. Ilkka Henrik M\"akinen and Prof. Olle Eriksson for support and advice.
This work was supported by the Swedish strategic research programme eSSENCE.

\bibliography{PRE-GenSirm}

\begin{thebibliography}{45}%
\makeatletter
\providecommand \@ifxundefined [1]{%
 \@ifx{#1\undefined}
}%
\providecommand \@ifnum [1]{%
 \ifnum #1\expandafter \@firstoftwo
 \else \expandafter \@secondoftwo
 \fi
}%
\providecommand \@ifx [1]{%
 \ifx #1\expandafter \@firstoftwo
 \else \expandafter \@secondoftwo
 \fi
}%
\providecommand \natexlab [1]{#1}%
\providecommand \enquote  [1]{``#1''}%
\providecommand \bibnamefont  [1]{#1}%
\providecommand \bibfnamefont [1]{#1}%
\providecommand \citenamefont [1]{#1}%
\providecommand \href@noop [0]{\@secondoftwo}%
\providecommand \href [0]{\begingroup \@sanitize@url \@href}%
\providecommand \@href[1]{\@@startlink{#1}\@@href}%
\providecommand \@@href[1]{\endgroup#1\@@endlink}%
\providecommand \@sanitize@url [0]{\catcode `\\12\catcode `\$12\catcode
  `\&12\catcode `\#12\catcode `\^12\catcode `\_12\catcode `\%12\relax}%
\providecommand \@@startlink[1]{}%
\providecommand \@@endlink[0]{}%
\providecommand \url  [0]{\begingroup\@sanitize@url \@url }%
\providecommand \@url [1]{\endgroup\@href {#1}{\urlprefix }}%
\providecommand \urlprefix  [0]{URL }%
\providecommand \Eprint [0]{\href }%
\providecommand \doibase [0]{http://dx.doi.org/}%
\providecommand \selectlanguage [0]{\@gobble}%
\providecommand \bibinfo  [0]{\@secondoftwo}%
\providecommand \bibfield  [0]{\@secondoftwo}%
\providecommand \translation [1]{[#1]}%
\providecommand \BibitemOpen [0]{}%
\providecommand \bibitemStop [0]{}%
\providecommand \bibitemNoStop [0]{.\EOS\space}%
\providecommand \EOS [0]{\spacefactor3000\relax}%
\providecommand \BibitemShut  [1]{\csname bibitem#1\endcsname}%
\let\auto@bib@innerbib\@empty
%</preamble>
\bibitem [{\citenamefont {Castellano}\ \emph {et~al.}(2009)\citenamefont
  {Castellano}, \citenamefont {Fortunato},\ and\ \citenamefont
  {Loreto}}]{RevModPhys.81.591}%
  \BibitemOpen
  \bibfield  {author} {\bibinfo {author} {\bibfnamefont {C.}~\bibnamefont
  {Castellano}}, \bibinfo {author} {\bibfnamefont {S.}~\bibnamefont
  {Fortunato}}, \ and\ \bibinfo {author} {\bibfnamefont {V.}~\bibnamefont
  {Loreto}},\ }\href {\doibase 10.1103/RevModPhys.81.591} {\bibfield  {journal}
  {\bibinfo  {journal} {Rev. Mod. Phys.}\ }\textbf {\bibinfo {volume} {81}},\
  \bibinfo {pages} {591} (\bibinfo {year} {2009})}\BibitemShut {NoStop}%
\bibitem [{\citenamefont {Clifford}\ and\ \citenamefont
  {Sudbury}(1973)}]{doi:10.1093/biomet/60.3.581}%
  \BibitemOpen
  \bibfield  {author} {\bibinfo {author} {\bibfnamefont {P.}~\bibnamefont
  {Clifford}}\ and\ \bibinfo {author} {\bibfnamefont {A.}~\bibnamefont
  {Sudbury}},\ }\href {\doibase 10.1093/biomet/60.3.581} {\bibfield  {journal}
  {\bibinfo  {journal} {Biometrika}\ }\textbf {\bibinfo {volume} {60}},\
  \bibinfo {pages} {581} (\bibinfo {year} {1973})}\BibitemShut {NoStop}%
\bibitem [{\citenamefont {Holley}\ and\ \citenamefont
  {Liggett}(1975)}]{holley1975}%
  \BibitemOpen
  \bibfield  {author} {\bibinfo {author} {\bibfnamefont {R.~A.}\ \bibnamefont
  {Holley}}\ and\ \bibinfo {author} {\bibfnamefont {T.~M.}\ \bibnamefont
  {Liggett}},\ }\href {\doibase 10.1214/aop/1176996306} {\bibfield  {journal}
  {\bibinfo  {journal} {Ann. Probab.}\ }\textbf {\bibinfo {volume} {3}},\
  \bibinfo {pages} {643} (\bibinfo {year} {1975})}\BibitemShut {NoStop}%
\bibitem [{\citenamefont {Frachebourg}\ and\ \citenamefont
  {Krapivsky}(1996)}]{PhysRevE.53.R3009}%
  \BibitemOpen
  \bibfield  {author} {\bibinfo {author} {\bibfnamefont {L.}~\bibnamefont
  {Frachebourg}}\ and\ \bibinfo {author} {\bibfnamefont {P.~L.}\ \bibnamefont
  {Krapivsky}},\ }\href {\doibase 10.1103/PhysRevE.53.R3009} {\bibfield
  {journal} {\bibinfo  {journal} {Phys. Rev. E}\ }\textbf {\bibinfo {volume}
  {53}},\ \bibinfo {pages} {R3009} (\bibinfo {year} {1996})}\BibitemShut
  {NoStop}%
\bibitem [{\citenamefont {Krapivsky}(1992)}]{PhysRevA.45.1067}%
  \BibitemOpen
  \bibfield  {author} {\bibinfo {author} {\bibfnamefont {P.~L.}\ \bibnamefont
  {Krapivsky}},\ }\href {\doibase 10.1103/PhysRevA.45.1067} {\bibfield
  {journal} {\bibinfo  {journal} {Phys. Rev. A}\ }\textbf {\bibinfo {volume}
  {45}},\ \bibinfo {pages} {1067} (\bibinfo {year} {1992})}\BibitemShut
  {NoStop}%
\bibitem [{\citenamefont {Wu}\ \emph {et~al.}(2004)\citenamefont {Wu},
  \citenamefont {Huberman}, \citenamefont {Adamic},\ and\ \citenamefont
  {Tyler}}]{wu2004information}%
  \BibitemOpen
  \bibfield  {author} {\bibinfo {author} {\bibfnamefont {F.}~\bibnamefont
  {Wu}}, \bibinfo {author} {\bibfnamefont {B.~A.}\ \bibnamefont {Huberman}},
  \bibinfo {author} {\bibfnamefont {L.~A.}\ \bibnamefont {Adamic}}, \ and\
  \bibinfo {author} {\bibfnamefont {J.~R.}\ \bibnamefont {Tyler}},\ }\href@noop
  {} {\bibfield  {journal} {\bibinfo  {journal} {Physica A: Statistical
  Mechanics and its Applications}\ }\textbf {\bibinfo {volume} {337}},\
  \bibinfo {pages} {327} (\bibinfo {year} {2004})}\BibitemShut {NoStop}%
\bibitem [{\citenamefont {Suchecki}\ \emph {et~al.}(2005)\citenamefont
  {Suchecki}, \citenamefont {Egu\'{\i}luz},\ and\ \citenamefont
  {San~Miguel}}]{PhysRevE.72.036132}%
  \BibitemOpen
  \bibfield  {author} {\bibinfo {author} {\bibfnamefont {K.}~\bibnamefont
  {Suchecki}}, \bibinfo {author} {\bibfnamefont {V.~M.}\ \bibnamefont
  {Egu\'{\i}luz}}, \ and\ \bibinfo {author} {\bibfnamefont {M.}~\bibnamefont
  {San~Miguel}},\ }\href {\doibase 10.1103/PhysRevE.72.036132} {\bibfield
  {journal} {\bibinfo  {journal} {Phys. Rev. E}\ }\textbf {\bibinfo {volume}
  {72}},\ \bibinfo {pages} {036132} (\bibinfo {year} {2005})}\BibitemShut
  {NoStop}%
\bibitem [{\citenamefont {Sood}\ and\ \citenamefont
  {Redner}(2005)}]{PhysRevLett.94.178701}%
  \BibitemOpen
  \bibfield  {author} {\bibinfo {author} {\bibfnamefont {V.}~\bibnamefont
  {Sood}}\ and\ \bibinfo {author} {\bibfnamefont {S.}~\bibnamefont {Redner}},\
  }\href {\doibase 10.1103/PhysRevLett.94.178701} {\bibfield  {journal}
  {\bibinfo  {journal} {Phys. Rev. Lett.}\ }\textbf {\bibinfo {volume} {94}},\
  \bibinfo {pages} {178701} (\bibinfo {year} {2005})}\BibitemShut {NoStop}%
\bibitem [{\citenamefont {Sood}\ \emph {et~al.}(2008)\citenamefont {Sood},
  \citenamefont {Antal},\ and\ \citenamefont {Redner}}]{sood2008voter}%
  \BibitemOpen
  \bibfield  {author} {\bibinfo {author} {\bibfnamefont {V.}~\bibnamefont
  {Sood}}, \bibinfo {author} {\bibfnamefont {T.}~\bibnamefont {Antal}}, \ and\
  \bibinfo {author} {\bibfnamefont {S.}~\bibnamefont {Redner}},\ }\href@noop {}
  {\bibfield  {journal} {\bibinfo  {journal} {Physical Review E}\ }\textbf
  {\bibinfo {volume} {77}},\ \bibinfo {pages} {041121} (\bibinfo {year}
  {2008})}\BibitemShut {NoStop}%
\bibitem [{\citenamefont {Sire}\ and\ \citenamefont
  {Majumdar}(1995)}]{sire1995coarsening}%
  \BibitemOpen
  \bibfield  {author} {\bibinfo {author} {\bibfnamefont {C.}~\bibnamefont
  {Sire}}\ and\ \bibinfo {author} {\bibfnamefont {S.~N.}\ \bibnamefont
  {Majumdar}},\ }\href@noop {} {\bibfield  {journal} {\bibinfo  {journal}
  {Physical Review E}\ }\textbf {\bibinfo {volume} {52}},\ \bibinfo {pages}
  {244} (\bibinfo {year} {1995})}\BibitemShut {NoStop}%
\bibitem [{\citenamefont {Mobilia}(2003)}]{PhysRevLett.91.028701}%
  \BibitemOpen
  \bibfield  {author} {\bibinfo {author} {\bibfnamefont {M.}~\bibnamefont
  {Mobilia}},\ }\href {\doibase 10.1103/PhysRevLett.91.028701} {\bibfield
  {journal} {\bibinfo  {journal} {Phys. Rev. Lett.}\ }\textbf {\bibinfo
  {volume} {91}},\ \bibinfo {pages} {028701} (\bibinfo {year}
  {2003})}\BibitemShut {NoStop}%
\bibitem [{\citenamefont {Mobilia}\ \emph {et~al.}(2007)\citenamefont
  {Mobilia}, \citenamefont {Petersen},\ and\ \citenamefont
  {Redner}}]{1742-5468-2007-08-P08029}%
  \BibitemOpen
  \bibfield  {author} {\bibinfo {author} {\bibfnamefont {M.}~\bibnamefont
  {Mobilia}}, \bibinfo {author} {\bibfnamefont {A.}~\bibnamefont {Petersen}}, \
  and\ \bibinfo {author} {\bibfnamefont {S.}~\bibnamefont {Redner}},\ }\href
  {http://stacks.iop.org/1742-5468/2007/i=08/a=P08029} {\bibfield  {journal}
  {\bibinfo  {journal} {Journal of Statistical Mechanics: Theory and
  Experiment}\ }\textbf {\bibinfo {volume} {2007}},\ \bibinfo {pages} {P08029}
  (\bibinfo {year} {2007})}\BibitemShut {NoStop}%
\bibitem [{\citenamefont {Vazquez}\ \emph {et~al.}(2003)\citenamefont
  {Vazquez}, \citenamefont {Krapivsky},\ and\ \citenamefont
  {Redner}}]{vazquez2003constrained}%
  \BibitemOpen
  \bibfield  {author} {\bibinfo {author} {\bibfnamefont {F.}~\bibnamefont
  {Vazquez}}, \bibinfo {author} {\bibfnamefont {P.~L.}\ \bibnamefont
  {Krapivsky}}, \ and\ \bibinfo {author} {\bibfnamefont {S.}~\bibnamefont
  {Redner}},\ }\href@noop {} {\bibfield  {journal} {\bibinfo  {journal}
  {Journal of Physics A: Mathematical and General}\ }\textbf {\bibinfo {volume}
  {36}},\ \bibinfo {pages} {L61} (\bibinfo {year} {2003})}\BibitemShut
  {NoStop}%
\bibitem [{\citenamefont {Lambiotte}\ and\ \citenamefont
  {Redner}(2007)}]{lambiotte2007dynamics}%
  \BibitemOpen
  \bibfield  {author} {\bibinfo {author} {\bibfnamefont {R.}~\bibnamefont
  {Lambiotte}}\ and\ \bibinfo {author} {\bibfnamefont {S.}~\bibnamefont
  {Redner}},\ }\href@noop {} {\bibfield  {journal} {\bibinfo  {journal}
  {Journal of Statistical Mechanics: Theory and Experiment}\ }\textbf {\bibinfo
  {volume} {2007}},\ \bibinfo {pages} {L10001} (\bibinfo {year}
  {2007})}\BibitemShut {NoStop}%
\bibitem [{\citenamefont {Dall'Asta}\ and\ \citenamefont
  {Castellano}(2007)}]{dall2007effective}%
  \BibitemOpen
  \bibfield  {author} {\bibinfo {author} {\bibfnamefont {L.}~\bibnamefont
  {Dall'Asta}}\ and\ \bibinfo {author} {\bibfnamefont {C.}~\bibnamefont
  {Castellano}},\ }\href@noop {} {\bibfield  {journal} {\bibinfo  {journal}
  {EPL (Europhysics Letters)}\ }\textbf {\bibinfo {volume} {77}},\ \bibinfo
  {pages} {60005} (\bibinfo {year} {2007})}\BibitemShut {NoStop}%
\bibitem [{\citenamefont {Fern\'andez-Gracia}\ \emph
  {et~al.}(2014)\citenamefont {Fern\'andez-Gracia}, \citenamefont {Suchecki},
  \citenamefont {Ramasco}, \citenamefont {San~Miguel},\ and\ \citenamefont
  {Egu\'{\i}luz}}]{PhysRevLett.112.158701}%
  \BibitemOpen
  \bibfield  {author} {\bibinfo {author} {\bibfnamefont {J.}~\bibnamefont
  {Fern\'andez-Gracia}}, \bibinfo {author} {\bibfnamefont {K.}~\bibnamefont
  {Suchecki}}, \bibinfo {author} {\bibfnamefont {J.~J.}\ \bibnamefont
  {Ramasco}}, \bibinfo {author} {\bibfnamefont {M.}~\bibnamefont {San~Miguel}},
  \ and\ \bibinfo {author} {\bibfnamefont {V.~M.}\ \bibnamefont
  {Egu\'{\i}luz}},\ }\href {\doibase 10.1103/PhysRevLett.112.158701} {\bibfield
   {journal} {\bibinfo  {journal} {Phys. Rev. Lett.}\ }\textbf {\bibinfo
  {volume} {112}},\ \bibinfo {pages} {158701} (\bibinfo {year}
  {2014})}\BibitemShut {NoStop}%
\bibitem [{\citenamefont {Fortunato}\ and\ \citenamefont
  {Castellano}(2007)}]{PhysRevLett.99.138701}%
  \BibitemOpen
  \bibfield  {author} {\bibinfo {author} {\bibfnamefont {S.}~\bibnamefont
  {Fortunato}}\ and\ \bibinfo {author} {\bibfnamefont {C.}~\bibnamefont
  {Castellano}},\ }\href {\doibase 10.1103/PhysRevLett.99.138701} {\bibfield
  {journal} {\bibinfo  {journal} {Phys. Rev. Lett.}\ }\textbf {\bibinfo
  {volume} {99}},\ \bibinfo {pages} {138701} (\bibinfo {year}
  {2007})}\BibitemShut {NoStop}%
\bibitem [{\citenamefont {Chatterjee}\ \emph {et~al.}(2013)\citenamefont
  {Chatterjee}, \citenamefont {Mitrovi{\'c}},\ and\ \citenamefont
  {Fortunato}}]{chatterjee2013universality}%
  \BibitemOpen
  \bibfield  {author} {\bibinfo {author} {\bibfnamefont {A.}~\bibnamefont
  {Chatterjee}}, \bibinfo {author} {\bibfnamefont {M.}~\bibnamefont
  {Mitrovi{\'c}}}, \ and\ \bibinfo {author} {\bibfnamefont {S.}~\bibnamefont
  {Fortunato}},\ }\href@noop {} {\bibfield  {journal} {\bibinfo  {journal}
  {Scientific reports}\ }\textbf {\bibinfo {volume} {3}},\ \bibinfo {pages}
  {1049} (\bibinfo {year} {2013})}\BibitemShut {NoStop}%
\bibitem [{\citenamefont {Klimek}\ \emph {et~al.}(2012)\citenamefont {Klimek},
  \citenamefont {Yegorov}, \citenamefont {Hanel},\ and\ \citenamefont
  {Thurner}}]{klimek2012statistical}%
  \BibitemOpen
  \bibfield  {author} {\bibinfo {author} {\bibfnamefont {P.}~\bibnamefont
  {Klimek}}, \bibinfo {author} {\bibfnamefont {Y.}~\bibnamefont {Yegorov}},
  \bibinfo {author} {\bibfnamefont {R.}~\bibnamefont {Hanel}}, \ and\ \bibinfo
  {author} {\bibfnamefont {S.}~\bibnamefont {Thurner}},\ }\href@noop {}
  {\bibfield  {journal} {\bibinfo  {journal} {Proceedings of the National
  Academy of Sciences}\ }\textbf {\bibinfo {volume} {109}},\ \bibinfo {pages}
  {16469} (\bibinfo {year} {2012})}\BibitemShut {NoStop}%
\bibitem [{\citenamefont {Borghesi}\ and\ \citenamefont
  {Bouchaud}(2010)}]{Borghesi2010}%
  \BibitemOpen
  \bibfield  {author} {\bibinfo {author} {\bibfnamefont {C.}~\bibnamefont
  {Borghesi}}\ and\ \bibinfo {author} {\bibfnamefont {J.-P.}\ \bibnamefont
  {Bouchaud}},\ }\href {\doibase 10.1140/epjb/e2010-00151-1} {\bibfield
  {journal} {\bibinfo  {journal} {The European Physical Journal B}\ }\textbf
  {\bibinfo {volume} {75}},\ \bibinfo {pages} {395} (\bibinfo {year}
  {2010})}\BibitemShut {NoStop}%
\bibitem [{\citenamefont {Borghesi}\ \emph {et~al.}(2012)\citenamefont
  {Borghesi}, \citenamefont {Raynal},\ and\ \citenamefont
  {Bouchaud}}]{borghesi2012election}%
  \BibitemOpen
  \bibfield  {author} {\bibinfo {author} {\bibfnamefont {C.}~\bibnamefont
  {Borghesi}}, \bibinfo {author} {\bibfnamefont {J.-C.}\ \bibnamefont
  {Raynal}}, \ and\ \bibinfo {author} {\bibfnamefont {J.-P.}\ \bibnamefont
  {Bouchaud}},\ }\href@noop {} {\bibfield  {journal} {\bibinfo  {journal} {PloS
  one}\ }\textbf {\bibinfo {volume} {7}},\ \bibinfo {pages} {e36289} (\bibinfo
  {year} {2012})}\BibitemShut {NoStop}%
\bibitem [{\citenamefont {Enikolopov}\ \emph {et~al.}(2013)\citenamefont
  {Enikolopov}, \citenamefont {Korovkin}, \citenamefont {Petrova},
  \citenamefont {Sonin},\ and\ \citenamefont {Zakharov}}]{enikolopov2013field}%
  \BibitemOpen
  \bibfield  {author} {\bibinfo {author} {\bibfnamefont {R.}~\bibnamefont
  {Enikolopov}}, \bibinfo {author} {\bibfnamefont {V.}~\bibnamefont
  {Korovkin}}, \bibinfo {author} {\bibfnamefont {M.}~\bibnamefont {Petrova}},
  \bibinfo {author} {\bibfnamefont {K.}~\bibnamefont {Sonin}}, \ and\ \bibinfo
  {author} {\bibfnamefont {A.}~\bibnamefont {Zakharov}},\ }\href@noop {}
  {\bibfield  {journal} {\bibinfo  {journal} {Proceedings of the National
  Academy of Sciences}\ }\textbf {\bibinfo {volume} {110}},\ \bibinfo {pages}
  {448} (\bibinfo {year} {2013})}\BibitemShut {NoStop}%
\bibitem [{\citenamefont {Filho}\ \emph {et~al.}(1999)\citenamefont {Filho},
  \citenamefont {Almeida}, \citenamefont {Andrade},\ and\ \citenamefont
  {Moreira}}]{PhysRevE.60.1067}%
  \BibitemOpen
  \bibfield  {author} {\bibinfo {author} {\bibfnamefont {R.~N.~C.}\
  \bibnamefont {Filho}}, \bibinfo {author} {\bibfnamefont {M.~P.}\ \bibnamefont
  {Almeida}}, \bibinfo {author} {\bibfnamefont {J.~S.}\ \bibnamefont
  {Andrade}}, \ and\ \bibinfo {author} {\bibfnamefont {J.~E.}\ \bibnamefont
  {Moreira}},\ }\href {\doibase 10.1103/PhysRevE.60.1067} {\bibfield  {journal}
  {\bibinfo  {journal} {Phys. Rev. E}\ }\textbf {\bibinfo {volume} {60}},\
  \bibinfo {pages} {1067} (\bibinfo {year} {1999})}\BibitemShut {NoStop}%
\bibitem [{\citenamefont {Araripe}\ and\ \citenamefont
  {Filho}(2009)}]{ARARIPE20094167}%
  \BibitemOpen
  \bibfield  {author} {\bibinfo {author} {\bibfnamefont {L.}~\bibnamefont
  {Araripe}}\ and\ \bibinfo {author} {\bibfnamefont {R.~C.}\ \bibnamefont
  {Filho}},\ }\href {\doibase https://doi.org/10.1016/j.physa.2009.06.023}
  {\bibfield  {journal} {\bibinfo  {journal} {Physica A: Statistical Mechanics
  and its Applications}\ }\textbf {\bibinfo {volume} {388}},\ \bibinfo {pages}
  {4167 } (\bibinfo {year} {2009})}\BibitemShut {NoStop}%
\bibitem [{\citenamefont {Bernardes}\ \emph {et~al.}(2002)\citenamefont
  {Bernardes}, \citenamefont {Stauffer},\ and\ \citenamefont
  {Kert{\'e}sz}}]{Bernardes2002}%
  \BibitemOpen
  \bibfield  {author} {\bibinfo {author} {\bibfnamefont {A.}~\bibnamefont
  {Bernardes}}, \bibinfo {author} {\bibfnamefont {D.}~\bibnamefont {Stauffer}},
  \ and\ \bibinfo {author} {\bibfnamefont {J.}~\bibnamefont {Kert{\'e}sz}},\
  }\href {\doibase 10.1140/e10051-002-0013-y} {\bibfield  {journal} {\bibinfo
  {journal} {The European Physical Journal B - Condensed Matter and Complex
  Systems}\ }\textbf {\bibinfo {volume} {25}},\ \bibinfo {pages} {123}
  (\bibinfo {year} {2002})}\BibitemShut {NoStop}%
\bibitem [{\citenamefont {Lyra}\ \emph {et~al.}(2003)\citenamefont {Lyra},
  \citenamefont {Costa}, \citenamefont {Costa~Filho},\ and\ \citenamefont
  {Andrade~Jr}}]{lyra2003generalized}%
  \BibitemOpen
  \bibfield  {author} {\bibinfo {author} {\bibfnamefont {M.}~\bibnamefont
  {Lyra}}, \bibinfo {author} {\bibfnamefont {U.}~\bibnamefont {Costa}},
  \bibinfo {author} {\bibfnamefont {R.}~\bibnamefont {Costa~Filho}}, \ and\
  \bibinfo {author} {\bibfnamefont {J.}~\bibnamefont {Andrade~Jr}},\
  }\href@noop {} {\bibfield  {journal} {\bibinfo  {journal} {EPL (Europhysics
  Letters)}\ }\textbf {\bibinfo {volume} {62}},\ \bibinfo {pages} {131}
  (\bibinfo {year} {2003})}\BibitemShut {NoStop}%
\bibitem [{\citenamefont {Travieso}\ and\ \citenamefont
  {da~Fontoura~Costa}(2006)}]{PhysRevE.74.036112}%
  \BibitemOpen
  \bibfield  {author} {\bibinfo {author} {\bibfnamefont {G.}~\bibnamefont
  {Travieso}}\ and\ \bibinfo {author} {\bibfnamefont {L.}~\bibnamefont
  {da~Fontoura~Costa}},\ }\href {\doibase 10.1103/PhysRevE.74.036112}
  {\bibfield  {journal} {\bibinfo  {journal} {Phys. Rev. E}\ }\textbf {\bibinfo
  {volume} {74}},\ \bibinfo {pages} {036112} (\bibinfo {year}
  {2006})}\BibitemShut {NoStop}%
\bibitem [{\citenamefont {Araripe}\ \emph {et~al.}(2006)\citenamefont
  {Araripe}, \citenamefont {Costa~Filho}, \citenamefont {Herrmann},\ and\
  \citenamefont {Andrade~Jr}}]{araripe2006plurality}%
  \BibitemOpen
  \bibfield  {author} {\bibinfo {author} {\bibfnamefont {L.~E.}\ \bibnamefont
  {Araripe}}, \bibinfo {author} {\bibfnamefont {R.~N.}\ \bibnamefont
  {Costa~Filho}}, \bibinfo {author} {\bibfnamefont {H.~J.}\ \bibnamefont
  {Herrmann}}, \ and\ \bibinfo {author} {\bibfnamefont {J.~S.}\ \bibnamefont
  {Andrade~Jr}},\ }\href@noop {} {\bibfield  {journal} {\bibinfo  {journal}
  {International Journal of Modern Physics C}\ }\textbf {\bibinfo {volume}
  {17}},\ \bibinfo {pages} {1809} (\bibinfo {year} {2006})}\BibitemShut
  {NoStop}%
\bibitem [{\citenamefont {Gonzalez}\ \emph {et~al.}(2004)\citenamefont
  {Gonzalez}, \citenamefont {Sousa},\ and\ \citenamefont
  {Herrmann}}]{gonzalez2004opinion}%
  \BibitemOpen
  \bibfield  {author} {\bibinfo {author} {\bibfnamefont {M.}~\bibnamefont
  {Gonzalez}}, \bibinfo {author} {\bibfnamefont {A.}~\bibnamefont {Sousa}}, \
  and\ \bibinfo {author} {\bibfnamefont {H.}~\bibnamefont {Herrmann}},\
  }\href@noop {} {\bibfield  {journal} {\bibinfo  {journal} {International
  journal of modern physics C}\ }\textbf {\bibinfo {volume} {15}},\ \bibinfo
  {pages} {45} (\bibinfo {year} {2004})}\BibitemShut {NoStop}%
\bibitem [{\citenamefont {Andresen}\ \emph {et~al.}(2008)\citenamefont
  {Andresen}, \citenamefont {Hansen}, \citenamefont {Hansen}, \citenamefont
  {Vasconcelos},\ and\ \citenamefont {Andrade~Jr}}]{andresen2008correlations}%
  \BibitemOpen
  \bibfield  {author} {\bibinfo {author} {\bibfnamefont {C.~A.}\ \bibnamefont
  {Andresen}}, \bibinfo {author} {\bibfnamefont {H.~F.}\ \bibnamefont
  {Hansen}}, \bibinfo {author} {\bibfnamefont {A.}~\bibnamefont {Hansen}},
  \bibinfo {author} {\bibfnamefont {G.~L.}\ \bibnamefont {Vasconcelos}}, \ and\
  \bibinfo {author} {\bibfnamefont {J.~S.}\ \bibnamefont {Andrade~Jr}},\
  }\href@noop {} {\bibfield  {journal} {\bibinfo  {journal} {International
  Journal of Modern Physics C}\ }\textbf {\bibinfo {volume} {19}},\ \bibinfo
  {pages} {1647} (\bibinfo {year} {2008})}\BibitemShut {NoStop}%
\bibitem [{\citenamefont
  {{Hern\`andez-Salda\~na}}(2009)}]{HERNANDEZSALDANA20092699}%
  \BibitemOpen
  \bibfield  {author} {\bibinfo {author} {\bibfnamefont {H.}~\bibnamefont
  {{Hern\`andez-Salda\~na}}},\ }\href {\doibase
  https://doi.org/10.1016/j.physa.2009.03.016} {\bibfield  {journal} {\bibinfo
  {journal} {Physica A: Statistical Mechanics and its Applications}\ }\textbf
  {\bibinfo {volume} {388}},\ \bibinfo {pages} {2699 } (\bibinfo {year}
  {2009})}\BibitemShut {NoStop}%
\bibitem [{\citenamefont {Ara{\'u}jo}\ \emph {et~al.}(2010)\citenamefont
  {Ara{\'u}jo}, \citenamefont {Andrade~Jr},\ and\ \citenamefont
  {Herrmann}}]{araujo2010tactical}%
  \BibitemOpen
  \bibfield  {author} {\bibinfo {author} {\bibfnamefont {N.~A.}\ \bibnamefont
  {Ara{\'u}jo}}, \bibinfo {author} {\bibfnamefont {J.~S.}\ \bibnamefont
  {Andrade~Jr}}, \ and\ \bibinfo {author} {\bibfnamefont {H.~J.}\ \bibnamefont
  {Herrmann}},\ }\href@noop {} {\bibfield  {journal} {\bibinfo  {journal} {PLoS
  One}\ }\textbf {\bibinfo {volume} {5}},\ \bibinfo {pages} {e12446} (\bibinfo
  {year} {2010})}\BibitemShut {NoStop}%
\bibitem [{\citenamefont {Kim}\ \emph {et~al.}(2003)\citenamefont {Kim},
  \citenamefont {Elliott},\ and\ \citenamefont {Wang}}]{KIM2003741}%
  \BibitemOpen
  \bibfield  {author} {\bibinfo {author} {\bibfnamefont {J.}~\bibnamefont
  {Kim}}, \bibinfo {author} {\bibfnamefont {E.}~\bibnamefont {Elliott}}, \ and\
  \bibinfo {author} {\bibfnamefont {D.-M.}\ \bibnamefont {Wang}},\ }\href
  {\doibase https://doi.org/10.1016/S0261-3794(02)00008-2} {\bibfield
  {journal} {\bibinfo  {journal} {Electoral Studies}\ }\textbf {\bibinfo
  {volume} {22}},\ \bibinfo {pages} {741 } (\bibinfo {year}
  {2003})}\BibitemShut {NoStop}%
\bibitem [{\citenamefont {Gardiner}(2009)}]{gardiner2009stochastic}%
  \BibitemOpen
  \bibfield  {author} {\bibinfo {author} {\bibfnamefont {C.}~\bibnamefont
  {Gardiner}},\ }\href@noop {} {\emph {\bibinfo {title} {Stochastic methods: A
  handbook for the Natural and Social Sciences}}},\ \bibinfo {edition} {fourth
  edition}\ ed.\ (\bibinfo  {publisher} {Springer Berlin},\ \bibinfo {year}
  {2009})\BibitemShut {NoStop}%
\bibitem [{\citenamefont {Scheucher}\ and\ \citenamefont
  {Spohn}(1988)}]{scheucher1988soluble}%
  \BibitemOpen
  \bibfield  {author} {\bibinfo {author} {\bibfnamefont {M.}~\bibnamefont
  {Scheucher}}\ and\ \bibinfo {author} {\bibfnamefont {H.}~\bibnamefont
  {Spohn}},\ }\href@noop {} {\bibfield  {journal} {\bibinfo  {journal} {Journal
  of statistical physics}\ }\textbf {\bibinfo {volume} {53}},\ \bibinfo {pages}
  {279} (\bibinfo {year} {1988})}\BibitemShut {NoStop}%
\bibitem [{\citenamefont {Granovsky}\ and\ \citenamefont
  {Madras}(1995)}]{granovsky1995noisy}%
  \BibitemOpen
  \bibfield  {author} {\bibinfo {author} {\bibfnamefont {B.~L.}\ \bibnamefont
  {Granovsky}}\ and\ \bibinfo {author} {\bibfnamefont {N.}~\bibnamefont
  {Madras}},\ }\href@noop {} {\bibfield  {journal} {\bibinfo  {journal}
  {Stochastic Processes and their applications}\ }\textbf {\bibinfo {volume}
  {55}},\ \bibinfo {pages} {23} (\bibinfo {year} {1995})}\BibitemShut {NoStop}%
\bibitem [{\citenamefont {Feld}(1981)}]{feld1981focused}%
  \BibitemOpen
  \bibfield  {author} {\bibinfo {author} {\bibfnamefont {S.~L.}\ \bibnamefont
  {Feld}},\ }\href@noop {} {\bibfield  {journal} {\bibinfo  {journal} {American
  journal of sociology}\ }\textbf {\bibinfo {volume} {86}},\ \bibinfo {pages}
  {1015} (\bibinfo {year} {1981})}\BibitemShut {NoStop}%
\bibitem [{\citenamefont {Feld}\ and\ \citenamefont
  {Grofman}(2009)}]{feld2009homophily}%
  \BibitemOpen
  \bibfield  {author} {\bibinfo {author} {\bibfnamefont {S.}~\bibnamefont
  {Feld}}\ and\ \bibinfo {author} {\bibfnamefont {B.}~\bibnamefont {Grofman}},\
  }\href@noop {} {\bibfield  {journal} {\bibinfo  {journal} {The Oxford
  handbook of analytical sociology}\ ,\ \bibinfo {pages} {521}} (\bibinfo
  {year} {2009})}\BibitemShut {NoStop}%
\bibitem [{\citenamefont {Michaud}(2017)}]{PhysRevE.95.022308}%
  \BibitemOpen
  \bibfield  {author} {\bibinfo {author} {\bibfnamefont {J.}~\bibnamefont
  {Michaud}},\ }\href {\doibase 10.1103/PhysRevE.95.022308} {\bibfield
  {journal} {\bibinfo  {journal} {Phys. Rev. E}\ }\textbf {\bibinfo {volume}
  {95}},\ \bibinfo {pages} {022308} (\bibinfo {year} {2017})}\BibitemShut
  {NoStop}%
\bibitem [{\citenamefont {Hart}\ and\ \citenamefont
  {Kurz}(1983)}]{hart1983endogenous}%
  \BibitemOpen
  \bibfield  {author} {\bibinfo {author} {\bibfnamefont {S.}~\bibnamefont
  {Hart}}\ and\ \bibinfo {author} {\bibfnamefont {M.}~\bibnamefont {Kurz}},\
  }\href@noop {} {\bibfield  {journal} {\bibinfo  {journal} {Econometrica:
  Journal of the Econometric Society}\ ,\ \bibinfo {pages} {1047}} (\bibinfo
  {year} {1983})}\BibitemShut {NoStop}%
\bibitem [{Note1()}]{Note1}%
  \BibitemOpen
  \bibinfo {note} {The code can be downloaded from \protect \href
  {https://www.researchgate.net/publication/322721295_Supplemental_material_for_Social_Influence_with_Recurrent_Mobility_with_multiple_options
  }{here.}}\BibitemShut {Stop}%
\bibitem [{Note2()}]{Note2}%
  \BibitemOpen
  \bibinfo {note} {In real election systems, like proportional elections, the
  voting dynamics is more complex and further work should be done to adapt our
  model to that case, see, for example, \cite
  {PhysRevLett.99.138701}.}\BibitemShut {Stop}%
\bibitem [{Note3()}]{Note3}%
  \BibitemOpen
  \bibinfo {note} {Opinions and parties are not fully independent of one
  another and opinion structures should be introduced at some point. In any
  VM-like model the opinion structure is ignored and opinion change only
  depends on demographics.}\BibitemShut {Stop}%
\bibitem [{\citenamefont {Hegselmann}\ and\ \citenamefont
  {Krause}(2002)}]{HegselmannKrause02}%
  \BibitemOpen
  \bibfield  {author} {\bibinfo {author} {\bibfnamefont {R.}~\bibnamefont
  {Hegselmann}}\ and\ \bibinfo {author} {\bibfnamefont {U.}~\bibnamefont
  {Krause}},\ }\href@noop {} {\bibfield  {journal} {\bibinfo  {journal}
  {Journal of Artificial Societies and Social Simulation}\ }\textbf {\bibinfo
  {volume} {5}} (\bibinfo {year} {2002})}\BibitemShut {NoStop}%
\bibitem [{\citenamefont {Axelrod}(1997)}]{doi:10.1177/0022002797041002001}%
  \BibitemOpen
  \bibfield  {author} {\bibinfo {author} {\bibfnamefont {R.}~\bibnamefont
  {Axelrod}},\ }\href {\doibase 10.1177/0022002797041002001} {\bibfield
  {journal} {\bibinfo  {journal} {Journal of Conflict Resolution}\ }\textbf
  {\bibinfo {volume} {41}},\ \bibinfo {pages} {203} (\bibinfo {year} {1997})},\
  \Eprint {http://arxiv.org/abs/https://doi.org/10.1177/0022002797041002001}
  {https://doi.org/10.1177/0022002797041002001} \BibitemShut {NoStop}%
\end{thebibliography}%
\end{document}